\documentclass[12pt]{iopart}

\usepackage{epsf,epsfig}
\usepackage{psfrag}
\usepackage{latexsym}
\usepackage{graphicx}

\newcommand{\beq}{\begin{equation}}
\newcommand{\eeq}{\end{equation}}

\newcommand{\ave}[1]{\mbox{$\langle #1\rangle$}}

\begin{document}

\title{Renewal processes and fluctuation analysis of molecular motor stepping}

\author{Jaime E. Santos\dag, Thomas Franosch\dag \ddag, Andrea
  Parmeggiani*, and Erwin Frey\ddag,\footnote[3]{To whom
    correspondence should be addressed (frey@lmu.de)}}

\address{\dag\ Hahn-Meitner Institut, Abteilung
  Theorie, Glienicker Str.100, D-14109 Berlin, Germany}

\address{\ddag\ Arnold Sommerfeld Center and CeNS, Department of
  Physics, Ludwig-Maximilians-Universit\"at M\"unchen,
  Theresienstrasse 37, D-80333 M\"unchen, Germany}

\address{*\ Laboratoire de Dynamique Mol\'eculaire des Interactions
  Membranaires, UMR 5539 CNRS/Universit\'e de Montpellier 2, Place
  Eug\`ene Bataillon, 34095 Montpellier Cedex 5, France.}

\date{\today}

\begin{abstract}
  We present a systematic method of analysis of experiments performed
  with single motors proteins. The use of such a method should allow 
  a more detailed description of the motor's chemical cycle through
  the precise fitting of the experimental data.
  We model the dynamics of a processive or rotary molecular motor
  using a renewal processes, in line with the work initiated by
  Svoboda, Mitra and Block.  We apply a functional technique to
  compute different types of multiple-time correlation functions of
  the renewal process, which have applications to bead-assay
  experiments performed both with processive molecular motors, such as
  myosin V and kinesin, and rotary motors, such as F1-ATPase.
\end{abstract}


\maketitle

\section{Introduction}
\label{secI}
There has been a growing interest in the biology and biophysics
communities in the study of the operation of motor
proteins \cite{Schliwa}. Such biological agents are typically single
enzymes that can act as thermodynamic engines, directly converting
chemical energy into mechanical energy \cite{fnote0} through a
chemical cycle which occurs at constant temperature \cite{Julicher99}.
Different classes of motor proteins are involved in biological
processes such as cellular transport, cellular mitosis or muscle
contraction, to cite a few examples of a plethora for which the
presence of such agents is crucial.

In this article, we concentrate on two particular
classes of motor proteins, namely linear processive molecular motors and
rotary molecular motors. Linear processive motor proteins, of which
myosin V and kinesin \cite{Rief00,Schief01} are important examples,
catalyse the ATP hydrolysis reaction, ATP $\rightarrow$ ADP+P, 
and use the energy so obtained to move along
linear molecular tracks, carrying organelles or membrane patches with
them and performing directed transport within cells with high
efficiency \cite{Alberts}. These organelles are too massive for
molecular diffusion to move them efficiently in a crowded environment
such as the cytoplasm within the time scales relevant for biological
processes.  The molecular tracks, which are composed of actin in the
case of myosin V and of tubulin in the case of kinesin, have a polar
character, i.e. such motors can move in only one direction. A different 
processive molecular motor, dynein, moves along the tubulin track in
the direction opposite to kinesin. 
Yet another example of a linear processive molecular motor is the
enzyme RNA-polymerase; see  \cite{Kafri04} and references therein for a 
more complete discussion of the characteristics of this motor. Here
it suffices to say that this enzyme moves along a DNA-strand,
also using ATP hydrolysis as its energy source. However, the function
of this enzyme is different from the motors described above, in that it does
not perform molecular transport; instead it promotes the transcription
of messenger-RNA from the underlying DNA-strand on which it moves.
Another significant difference is that the motion of this motor occurs
in an heterogeneous medium such as the DNA-strand, rather than in  
homogeneous media such as the actin or tubulin molecular tracks.
 
The chemical cycle that results
in the hydrolysis of one ATP molecule by the molecular motor is
composed of several substeps, corresponding to changes in the
internal conformation of the motor, accompanied by chemical reactions,
e.g. ATP binding to the motor, ATP hydrolysis, ADP and P release, etc.
Typically, one ATP hydrolysis is necessary for the motor to advance
one step along the track, docking to the next available site in it,
the size of such steps being around 35 nm in the case of myosin V and
8 nm in the case of kinesin  \cite{Schnitzer97,Mehta99,Ali02,fnote_Ali}. 
It should also
be noted that the use of the word 'processive' to classify a molecular
motor is limited to motors whose dwell time on the molecular track is
much larger than the time for a complete chemical cycle, i.e. the
typical time for a single step along the track.

Another class of molecular motors to which one may also apply 
some of the results presented in this paper are 
the so-called rotary molecular motors,
of which F1-ATPase is perhaps the most 
important example \cite{Yoshida01}. This motor
can also catalyse the ATP hydrolysis reaction and use the energy
so obtained to generate the rotary motion of a shaft-like mechanism. 
The mechano-chemical cycle of such a motor is composed of three rotations
of 120 degrees, determined by the symmetry of the molecule in question,
each rotation being coupled to the hydrolysis of one ATP molecule, as described
by the so-called ``binding change mechanism'' \cite{Boyer97}.
If such a motor is coupled to the proton-driven Fo motor, the assembly of
these two motors can either work as a proton pump or 
use a proton gradient to synthetise ATP, hence 
its name, ATPsynthase.

A great wealth of knowledge about the internal chemical cycle of
linear processive molecular motors has been acquired through
bead-assay experiments \cite{Svoboda94,Mehta_Review_99,
Mehta99,Mehta_Commentary,Rief00,Schief01,Veigel02,Block03}.
In such an experiment, performed in a fluid medium, a molecular motor
is coupled to a dielectric bead of micrometric size, the
size of such a sphere being nevertheless much larger than that 
of the motor protein.  The bead can be manipulated through the 
use of optical or
magnetic tweezers. In certain situations, one can, using an optical
trap, perform experiments in which the molecular motor, moving along
its track, is subjected to a known constant force of several pN. The
study of the motion for different values of the applied force allows
one to determine the force-velocity relation characteristic of a given
motor. At a given value of the force, known as the stalling force, the
motor usually decouples from the track (such decoupling can even occur
in the absence of force, but it is a much rarer event in that case
 \cite{Schnitzer00}) or it may move backwards along the track. Note
that such experiments may be performed at distinct values of the
applied load and also at different ATP concentrations.

More recently, Cappello et al. \cite{Cappello03,Badoual04} have
considered an experimental apparatus containing a bead-motor assay
that moves through the interference fringes of an evanescent
light-wave that exists in the proximity of a microscope glass plate.
The force which the electric field of such wave exerts on the bead is
negligible, i.e. the experiment is performed in a zero-load condition.
However, the bead still scatters photons of the evanescent field and
the observation of such events may be used to track the bead's position with
high spatial and time resolutions. In particular, such a method allows
for measurements with time resolutions of $\mu s$ (MHz). In normal
bead-assays, the feedback mechanism used to keep the bead
under constant force typically limits the time resolution to a few
milliseconds (kHz). Note that one may perform these measurements at
different ATP concentrations, but one is always limited to work at
zero external load, which constitutes both a strength (because it
permits higher time resolutions) and a weakness of the method (because
it limits the parameter range in which the system can be studied).

In essence, a clear qualitative picture has emerged from the different
types of bead-assay experiments, namely that a few of the substeps
which compose the chemical cycle of a molecular motor (known in the
literature as the rate-limiting steps) are of particularly long duration
compared to the remaining substeps, and that the statistics of the
motion of the molecular motor are, for the time re\-so\-lu\-tions
available, essentially determined by the duration of such
rate-limiting steps. Thus, the experimental study of the motion of
linear processive molecular motors can provide valuable information
concerning the nature and the duration of the rate-limiting steps
under different load conditions or under different concentrations of
ATP \cite{Block03}.  There is experimental evidence that two
rate-limiting steps are sufficient to describe the chemical cycle of
myosin V, namely ATP binding (by lowering enough the concentration of
ATP it is always possible to make ATP binding to the motor a
rate-limiting step) and ADP release \cite{Rief00}, whereas three or
four rate-limiting steps may be needed to describe the chemical cycle
of kinesin, depending on the ATP concentration. It is not yet well
understood to which chemical substeps these rates correspond \cite{Block03}.

The rotary movement of the $\gamma$ sub-unit of the
F1-ATPase motor can be visualised by
coupling such a motor to a microprobe, which can be either a
fluorescent actin filament, a gold or
polysterene bead, or a single fluorescent 
dye \cite{Yoshida01,Adachi00}. The rotation of the probe can be
recorded with a microscope and a CCD camera. It follows from such
studies that the 360-degree rotation can be decomposed into three 
120-degree substeps, each of which is in turn composed of two
rate-limiting steps with approximately the same duration. These steps
involve rotations of 90 and 30 degrees, respectively \cite{Kinosita04}.
One should note that of the three different types of microprobe
experiments referred, the use of 
a single fluorescent dye seems to be the most promising method of 
visualisation \cite{Yoshida01,Adachi00}, as it does not involve a 
perturbation of the motion
of the molecule. In the other cases, the rotation of F1-ATPase
is hindered by the large frictional coupling between the microprobe (actin
filament or gold or polysterene bead) and the surrounding fluid.
 
The theoretical tool that we will be applying to the analysis of such
experiments is the concept of renewal process.  Such processes are
ubiquitous in physics. These models have been successfully applied to
describe the motion of processive molecular motors such as myosin V
and kinesin \cite{Svoboda94,Cappello03,Badoual04}, the statistics of
detection of quantum particles \cite{QO,Benard73,Apana95} and
persistence phenomena in kinetic Ising models \cite{Godreche01,Allegrini04}.
Also, the results that relate to persistent phenomena have 
applications to studies
of the volatility of financial markets \cite{Giardina_PhysicaA}. The
above list is not exhaustive.

Loosely speaking, a renewal process is a counting process where unit
increments occur at random times.  We will consider only independently
distributed renewal processes, i.e. processes in which the probability
distribution for the occurrence of the next increment (also known as
the waiting-time distribution of the renewal process) depends only on
the time elapsed since the occurrence of the last increment and not on
the previous history of the counting process.

If one wishes to be more specific, one can say that the purpose of
this article is to enumerate and classify a series of 
bead-assay experiments performed with linear processive molecular motors or
with rotary molecular motors. These experiments are characterised by the
fact that one can apply the simple (and exact) results obtained
from the calculation of multiple-time correlation functions of renewal
processes with an arbitrary waiting-time distribution to the analysis
of the data obtained. 
In order to apply such simple models to linear processive motors, 
one represents a forward step of a molecular motor by an increment 
of the renewal
process, whose waiting-time distribution is dependent on the number of
rate-limiting steps of the motor. Such an approximation is justified
if the molecular motor performs backward steps only infrequently within
the time-window of observation, i.e. provided one is not working too
close to the stall force or at ATP concentrations which are too low,
when the rates for a forward or a backward step to occur become
comparable and one needs to use a large time-window of observation.
The application of these results to the dynamics of rotary motors
involves the mapping of a rotation of the motor by a given angle to an
increment of the renewal process, but such a mapping is more subtle
(see below for details). Such a mapping can be justified provided that
the motor performs backward rotations only infrequently  within the 
time-window of observation, which is the case if one is working at high ATP
concentrations.

The measurement of single-time correlation functions of the number of
increments $N(T)$, which have occurred in a renewal process until a given 
time $T$, is a common practice in the context of the experimental study 
of linear processive molecular motors.  The measurement of these 
quantities was first undertaken by Svoboda and coworkers in their 
experiments performed with the linear processive molecular motor 
kinesin \cite{Svoboda94}.  These
authors have considered the behaviour of the first and second moments
of $N(T)$, i.e. $\langle N(T)\rangle$ and $\langle N^2(T)\rangle$,
where the averaging is taken over different realisations of the
experiment. Such correlation functions contain information concerning
both the number of rate-limiting steps in a chemical cycle and the
characteristic rates pertaining to such steps.  
However, such single-time correlation functions do not fully characterise the
motor's chemical cycle, in particular in the case of motors whose cycles are
composed of many rate-limiting steps, like kinesin.  The measurement
of multiple-time correlation functions can provide additional valuable
information in such a case \cite{Cappello03,Badoual04}.

We employ a method based on the use of the probability-generating
functional \cite{Daley_Vere-Jones,Honerkamp} to compute multiple-time
correlation functions of a renewal process. For simplicity, and due to
its experimental relevance, we explicitly compute the mean-square 
deviation of the number of increments that occur between time $t_2$ 
and a later time $t_1$. This correlation function 
is  mathematically defined as $\langle (N(t_1)-N(t_2))^2
\rangle -\langle N(t_1)-N(t_2)\rangle^2$, 
where the times $t_2$ and $t_1$ are large 
compared to the typical time of a single chemical cycle.  We show that this 
function contains additional information concerning 
the rate-limiting steps of the chemical process, information
which cannot be extracted from single-time correlation functions. 
We also briefly indicate how the computation of higher-order correlation 
functions can be performed. Furthermore, we explicitly compute 
the density-density correlation function for a single motor, 
which was considered in the
experiments of Cappello and co-workers \cite{Cappello03,Badoual04}.
Interestingly enough, one can show that in a certain limiting case,
such a function is identical to the spin-spin correlation function
considered by Godreche and Luck \cite{Godreche01} in their
study of persistence phenomena in kinetic Ising models. Furthermore,
the class of spin models obtained in this particular limit
(to which the model studied by Godreche and Luck belongs) may be
relevant for the experimental study of rotary molecular motors, and
hence it is also discussed here.

The structure of this paper is as follows: in the next section, we
present the general results obtained and relate such results to the
relevant experiments. In section \ref{secIII}, we provide a
mathematical introduction to renewal processes,
following \cite{Daley_Vere-Jones} and introduce the probability-generating 
functional of the renewal process, together with some
related quantities whose usefullness will become apparent in section
\ref{secIV}. In section \ref{secIV}, we provide a general derivation of the
results presented in section \ref{secII}, concerning multiple-time
correlation functions of a renewal process, including the mean-square
deviation and the density-density correlation functions mentioned above, 
whose expressions we will explicitly compute in the asymptotic regime of
large times. A reader whose primary interest is not mathematics may
skip sections \ref{secIII} and \ref{secIV} 
without loss of continuity with respect to the
remainder of the paper. Finally, in section \ref{secV}, we present our
conclusions and provide a brief outlook of the experimental work
that we believe can be carried in this field using the results that
we have derived.  In the appendices, we discuss two simple examples of
waiting-time distributions for which the quantities discussed in the
main text can be computed outside the asymptotic regime.

\section{General discussion} 
\label{secII}
The most elementary quantity one can measure in a bead-assay
experiment is the average displacement of a motor $\langle
x(T)\rangle$ up to time $T$ over many runs. The simplest model 
for such a motor is that $x(T)=N(T)\,d$, where $N(T)$ is an
integer variable, updated at random intervals with a given
distribution $f(\zeta)$ (a renewal process), and $d$ is the motor's step
size. In the limit of large time, Blackwell's renewal
theorem \cite{Cox} guaranties that $\langle x(T)\rangle\ =
T\,d/\langle\tau\rangle$ asymptotically, where
$\ave{\tau}=\int_0^\infty\,d\zeta \,\zeta f(\zeta)$ is the 
average waiting time of
the distribution and corresponds to the mean duration time (or
turnover time) of a single chemical cycle.

The simplest choice one can take for $f(\zeta)$ is the exponential
distribution $f(\zeta)=e^{-\zeta/\tau}/\tau$. In this case, $\ave{\tau}=\tau$.
In their discussion of bead-motor assay experiments performed on the
processive motor kinesin, Svoboda, Mitra and Block \cite{Svoboda94}
have examined the more complicated situation where the waiting-time
distribution $f(\zeta)$ is given by the convolution of a finite number
${\cal M}$ of simple Poisson processes. Each of these processes, which
occur in series with typical times $\tau_1,\cdots,\tau_{\cal M}$, is
supposed to represent a rate-limiting step of the motor's chemical
cycle.  In such a case, the average waiting time is given by
\begin{equation}
\ave{\tau}=\sum_{i=1}^{\cal M}\tau_i\,. 
\label{GDeq0}
\end{equation}
Such a model is in good agreement with experiments where one measures
the distance travelled by the molecular motor alone. If one wishes to
resolve the chemical cycle substeps and its associated pathways, e.g.
through the use of cryoelectron microscopy \cite{Hoenger00,Schief01},
or by using bead-motor assays where lateral or forward loads are applied
to the motor \cite{Block03}, one needs to make use of more involved
models \cite{Juelicher97,Parmeggiani99,Schief01,Parmeggiani01,Lattanzi01,
Fisher01,Kolomeisky03,Lipowsky03,Vilfan05}
in order to interpret such experiments.

Svoboda et al. \cite{Svoboda94} have also introduced the concept of the
randomness coefficient $r$, characteristic of a single molecular
motor.  Given that the quantity $\langle x^2(T)\rangle - \langle
x(T)\rangle^2$ is the mean square deviation of the distance travelled
by the motor, $r$ is defined as
\begin{equation}
\label{GDeq1}
 r=\lim_{T\rightarrow\infty}\frac{\langle x^2(T)\rangle - \langle
x(T)\rangle^2}{\langle x(T)\rangle\,d}\,.
\end{equation}
The randomness coefficient, being the ratio of the mean-square
deviation of distance travelled by the motor to the average distance
travelled itself, represents a measure of the deviation of the motors
stepping from a deterministic motion, which would occur if $r=0$.
Also, note that $r$ is chosen to be dimensionless \cite{fnotePeclet}.
For the simple Poisson process, $\langle x^2(T)\rangle - \langle
x(T)\rangle^2=T\,d^2/\tau$, i.e. $r=1$.  For a renewal process
composed of ${\cal M}$ rate-limiting steps \cite{fnote_pedantic}, one
can show \cite{Svoboda94} that the randomness parameter is given by
\begin{equation}
\label{GDeq2}
 r=\frac{\sum_{i=1}^{\cal M}\tau_i^2}{
\left(\sum_{i=1}^{\cal M}\tau_i\right)^2}.
\end{equation}
It follows from (\ref{GDeq2}) that if ${\cal M}>1$, $r<1$.  For a
renewal process composed of ${\cal M}$ rate-limiting steps, $r\geq
1/{\cal M}$, the equality being obtained when all rates are equal.  In
the limit of an infinite number of substeps whose characteristic time
tends to zero, $r=0$ and the motor performs a deterministic motion.

The measurement of the randomness parameter, which is robust against
thermal noise \cite{Svoboda94} or the influence of the initial
conditions (i.e. of the experimental set up), is a powerful
experimental tool that can be used to rule out a proposed chemical
cycle, if such a cycle contains too small a number of rate-limiting
steps. For example, a measurement of $r<1/2$ indicates that 
at least three rate-limiting steps are needed to
describe the motor's chemical cycle \cite{fnote1}.

Since the chemical cycle of myosin V appears to be composed of only
two rate-limiting steps, the joint measurement of $r$ and of the
average distance travelled by the motor at large times is sufficient
to determine the value of the typical times $\tau_1$ and $\tau_2$
associated to each substep.  However, if one is considering the
experimental study of motors with more than two rate-limiting steps in
their cycle, such as kinesin, the measurement of these two quantities
is not equivalent to the complete determination of the characteristic
times $\tau_1,\dots,\tau_{\cal M}$ (${\cal M}>2$).

The statistical analysis of step duration using such bead-motors
assays, which allows for the direct extraction of the waiting-time
distribution $f(\zeta)$, was also considered by several
authors \cite{Mehta99,Rief00}. By fitting $f(\zeta)$ with an appropriate
convolution of exponential functions, one can obtain the values of all
the characteristic times $\tau_1,\dots,\tau_{\cal M}$ for arbitrary
${\cal M}$. However, such a technique requires the substitution of a
signal with a rich structure (the individual trajectories of the motor
as a function of time) by a function in steps, which leads to an
effective truncation of the data, because possible substeps within the
chemical cycle are erased.  Furthermore, such a technique requires a
resolution of individual steps, which is not always feasible.

One is thus led to consider the information provided by multiple-time
correlation functions.  The simplest quantity one can consider is the
average number of steps given by the motor between times $t_2$ and
$t_1$, i.e. the correlation function $\langle N(t_1)-N(t_2)\rangle$,
which for large times $t_1,t_2\rightarrow \infty$ behaves as
$(t_1-t_2)/\ave{\tau}$, and which obviously does not carry any new
information. Its measurement yields, as above, the value of the
turnover time $\ave{\tau}$ of the chemical cycle.

The same cannot be said of the mean-square deviation of this quantity,
i.e. the connected correlation function $\langle
(N(t_1)-N(t_2))^2\rangle_{\mbox{\scriptsize conn}}= \langle
(N(t_1)-N(t_2))^2\rangle -\langle N(t_1)-N(t_2)\rangle^2$.  If one
were to take $t_2=0$, this correlation function would reduce to the
mean square deviation of the distance travelled by the motor up to
time $t_1$, introduced above, and from which one can extract
the randomness parameter, as defined in (\ref{GDeq1}). However, there
are good reasons to consider instead the opposite limit in which both
$t_1, t_2\rightarrow \infty$, with $t=t_1-t_2$ kept finite, as in this
limit the correlation function becomes dependent only on $t$, i.e. one
recovers a form of time-translation invariance (we will write the
correlation function in this limit as ${\cal C}(t)=
\lim_{t_1,t_2\rightarrow \infty} \langle
(N(t_1)-N(t_2))^2\rangle_{\mbox{\scriptsize conn}}$). Such a limit is of
experimental relevance, since such a correlation function becomes
independent of the initial conditions, which can change from one
experimental realisation to the other.  Furthermore, its measurement
contains additional information concerning the waiting-time
distribution, rather than just the value of the randomness parameter,
as we will now discuss. Finally, such a measurement, just like the
measurement of the average number of steps $\langle
N(t_1)-N(t_2)\rangle$, requires only the ensemble averaging of the
data and is not plagued by the limitations one encounters if one tries
to measure $f(\zeta)$ directly, as described above.

In our treatment, we will consider a renewal process with an arbitrary
waiting-time distribution, rather than the special choice made by
Svoboda et al., since the study of the general case does not involve a
more complicated analysis.  However, we will always indicate the
results for this particular waiting-time distribution, given its
experimental importance.

It will be shown in section \ref{secIV} that this correlation function
displays the following behaviour in the limit of small and large time
differences $t=t_1-t_2$,
\begin{eqnarray}
\label{GDeq3}
{\cal C}(t)&=&
\left\{{ \frac{t}{\ave{\tau}}\;\;\mbox{if}\;\;t \ll \ave{\tau} \atop 
\frac{r}{\ave{\tau}}\,t+C\;\;\mbox{if}\;\;t\gg \ave{\tau}}
\right.,
\end{eqnarray}
with $r=\frac{\ave{\tau^2}-\ave{\tau}^2}{\ave{\tau}^2}$ and
$C=\frac{\ave{\tau^2}^2}{2\ave{\tau}^4}-
\frac{\ave{\tau^3}}{3\ave{\tau}^3}$, where
$\ave{\tau^2}=\int_0^\infty\,d\zeta \,\zeta^2 f(\zeta)$ and
$\ave{\tau^3}=\int_0^\infty\,d\zeta\,\zeta^3 f(\zeta)$. Please note that the
result just quoted for $r$, the randomness parameter, is a
generalisation of equation (\ref{GDeq2}) to the case of general
waiting-time distribution $f(\zeta)$ \cite{Godreche01}. The constant $C$
can be written in terms of $r$ and of the connected third-moment 
of the waiting-time distribution,
$\delta=\frac{\ave{(\tau-\ave{\tau})^3}}{2\ave{\tau}^3}$, as
\begin{equation}
C=\frac{1}{6}+\frac{1}{2}r^2-\frac{2}{3}\,\delta\,.
\label{GDeq3A}
\end{equation}
In the case in which $f(\zeta)$ is given by the convolution 
of ${\cal M}$ Poisson
processes, as considered by Svoboda et al., one can show (see section
\ref{secIV}) that $r$ reduces to equation (\ref{GDeq2}). In this case,
$\delta=\frac{\sum_{i=1}^{\cal M}\tau_i^3}{\left(\sum_{i=1}^{\cal M}
\tau_i\right)^3}$. The constant $C$ is exactly zero for a simple Poisson
process.

\begin{figure}[htbp]
\centerline{\epsfxsize 0.5\columnwidth \epsfbox{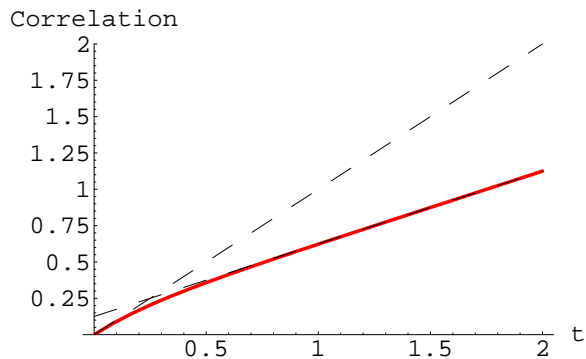}}
\caption{ Mean-square deviation of the number of steps in the
  regime of large times (red curve), for a motor with two rate-limiting steps.
  Also shown are the linear regimes at small and large time $t$.}
\label{F1}
\end{figure}

\begin{figure}[htbp]
\centerline{\epsfxsize 0.5\columnwidth \epsfbox{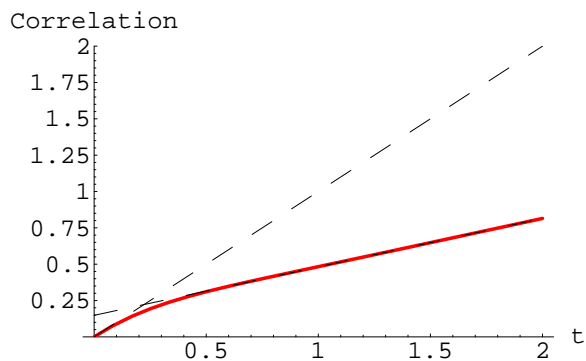}}
\caption{Mean-square deviation of the number of steps in the regime of
  large times (red curve), for a motor with three rate-limiting steps. 
  Also shown are the linear regimes at small and large time $t$.}
\label{F2}
\end{figure}

As we will show in \ref{apA}, one can compute ${\cal C}(t)$
for arbitrary values of $t$, rather than just the asymptotic limits
given by (\ref{GDeq3}), for processes whose waiting-time distribution
is given by the convolution of two or three Poisson processes, i.e. if
${\cal M}=2,3$ (the case ${\cal M}=1$ is trivial, see section
\ref{secIV}).  Such computations are explicitly performed because of
the importance of such processes for the experimental study of myosin
V and kinesin.  We plot the results of these calculations below, in
figures \ref{F1} and \ref{F2}, respectively, with the choices
$\tau_1=\tau_2=1/2$ for the case of a renewal process composed of two
rate-limiting steps and $\tau_1=\tau_2=\tau_3=1/3$ for the case of a
renewal process composed of three rate-limiting steps.  In both cases,
$\langle \tau\rangle =1$ (in arbitrary units), with $r=1/2$ in the
first case and $r=1/3$ in the second case. It is seen that the two
functions have the correct asymptotic limits at short and large times,
as given by equation (\ref{GDeq3}).

As stated above, the measurement of the average distance travelled by
the motor permits one to determine the turnover time $\ave{\tau}$. Once
this quantity is known, one can use the measurement of ${\cal C}(t)$
at large times to determine both the randomness coefficient $r$ (which
is given by the slope of the straight line multipled by $\ave{\tau}$),
as well as the constant $C$ which is related to the second and third
moment of $f(\zeta)$ \cite{fnote2}.  For a motor whose chemical cycle is
composed of two rate-limiting steps, the constant $C$ is given by
$C=\frac{2\tau_1^2\tau_2^2}{(\tau_1+\tau_2)^4}$. The measurement of
$C$ does not provide any new information in this case, but it may
provide a way to check or to improve the results obtained from the
joint measurement of the average distance travelled by the motor and
of the randomness parameter $r$. On the other hand, in the case of a
motor whose chemical cycle is composed of three rate-limiting steps,
the measurement of the total distance travelled by the motor, which
permits one to determine the total turnover time
$\langle\tau\rangle=\tau_1+\tau_2+\tau_3$, of $r$ and of
$C=\frac{2}{(\tau_1+\tau_2+\tau_3)^4}\,(\tau_1^2\tau_2^2+
\tau_1^2\tau_2\tau_3+\tau_1\tau_2^2\tau_3+\tau_1^2\tau_3^2+\tau_1\tau_2\tau_3^2
+\tau_2^2\tau_3^2)$ can be used to determine the three time-constants
$\tau_1$, $\tau_2$ and $\tau_3$, through a fit of the experimental
results.  Such a measurement may be particularly useful in experiments
with kinesin and for concentrations of ATP for which ATP binding is
not a rate-limiting step, because in such a case the chemical cycle of
kinesin appears to be composed of three rate-limiting
steps \cite{Block03}.

Another example of a multiple-time correlation function which can be
measured experimentally is the density-density correlator of an
ensemble of independent molecular motors, defined as
$S(q,t_1,t_2)=\langle\,e^{-iqd\,(N(t_1)-N(t_2))}\,\rangle$, where $q$
has the dimensions of a wave-vector and $d$ is the motor's step size.
This quantity is the Fourier transform of the probability for the
motor to move by a distance $n\times d$ between time $t_2$ and $t_1$.
It was directly measured in the experiments of Cappello and
coworkers \cite{fnote_CB}, where the bead-motor assay moves through the
interference fringes of an evanescent wave. In such an experiment,
$q=2\pi/\Lambda$, where $\Lambda$ is the period of the mask used to
create the interference pattern \cite{Cappello03,Badoual04}.  In the
long time limit $t_2,t_1\rightarrow \infty$, $S(q,t_1,t_2)$ becomes
solely dependent on the time difference $t=t_1-t_2$, in which case one
simply writes $S(q,t)$. One can show (see section \ref{secIV}) that
the Laplace transform of such a function,
$\tilde{S}(q,s)=\int_0^\infty\,dt\,e^{-st}\,S(q,t)$, is given, for a
general distribution $f(\zeta)$ with a finite average time
$\langle\tau\rangle$, by
\begin{equation}
\label{GDeq4}
\tilde{S}(q,s)=\frac{1}{s}\,\left(1+\frac{(e^{-iqd}-1)(1-\tilde{f}(s))}{
\langle\tau\rangle\,s\,(1-e^{-iqd}\tilde{f}(s))}\right)\,,
\end{equation}
where $\tilde{f}(s)$ is the Laplace transform of $f(\zeta)$. Note that in
the experiments of Cappello and co-workers, this function was taken to
be a simple exponential distribution. The equation above generalises
their result to the case of a general waiting-time distribution.

One can relate the Fourier transform of $S(q,t)$,
$S(q,\omega)=\int_{-\infty}^{+\infty}\,dt\,e^{i\omega t}\,S(q,t)$ to
its Laplace transform by $S(q,\omega)=2 Re\tilde{S}(q,s=-i\omega)$.
The Fourier transform is more amenable to computation from the
measured data than the Laplace transform, or can otherwise be directly
measured. One can show from such a relation and from equation
(\ref{GDeq4}) that $S(q,\omega)$ is given by
\begin{equation}
\label{GDeq5}
S(q,\omega)=
\frac{2(1-\cos(qd))(1-|\tilde{f}(-i\omega)|^2)}{\omega^2\,
\langle\tau\rangle\,|1-e^{-iqd}\tilde{f}(-i \omega)|^2}\,.
\end{equation}
If $f(\zeta)$ is a simple exponential distribution, one obtains from 
(\ref{GDeq5}),  
\begin{equation}
\label{GDeq6}
S(q,\omega)=\frac{2\tau(1-\cos(qd))}{(\omega\tau-\sin (qd))^2+(1-\cos(qd))^2}\,,
\end{equation}
which becomes the result for asymetric diffusion in the continuum
limit $qd \ll 1$, i.e. one obtains in this limit
\begin{equation}
\label{GDeq6a}
S(q,\omega)=\frac{2 D\,q^2}{(\omega-qv)^2+D^2\,q^4}\,,
\end{equation}
where we have introduced the motor's velocity $v=d/\tau$ and the motor's
diffusion constant $D=d^2/2\tau$, valid for a Poisson process, since
equation (\ref{GDeq6a}) was derived from (\ref{GDeq6}). For 
a renewal process with waiting-time distribution $f(\zeta)$, one can still
approximate $S(q,\omega)$ by (\ref{GDeq6a}) if the Laplace transform 
$\tilde{f}(s)$  is analytic at $s=0$, i.e. if all moments of the distribution 
$f(\zeta)$ exist.
In such a case, $v=d/\ave{\tau}$ and $D=d^2\,r/2\ave{\tau}$. 
Such approximation is valid at low-frequencies compared to the total
turnover rate and at low-momenta compared to the inverse step size,
i.e. if $\omega\ave{\tau}\ll 1$ and $qd\ll 1$. In other words, the
motor behaves in such a limit as a Brownian particle, characterized by
the two parameters $v$ and $D$ \cite{Badoual04}, as one would expect.
In \ref{apB}, we will explicitly compute $S(q,\omega)$
outside this asymptotic region for the cases in which $f(\zeta)$ is given
by the convolution of two or three Poisson processes, which
corresponds to a chemical cycle composed of two or three rate-limiting
steps. Again, such calculations are carried through because of the
experimental significance of these two cases for the study of myosin V
and of kinesin, since a measurement of $S(q,t)$ and the subsequent
computation of $S(q,\omega)$ would permit the extraction of the
relevant time constants by fitting the measured $S(q,\omega)$ with the
corresponding expression, appropriate for the given number of
rate-limiting steps, as given in \ref{apB} \cite{fnote_Cappello_Exp}. 
Note that in
order to perform such a fitting, one should first have a qualitative
understanding of the nature of the chemical cycle, in particular of
the number of rate-limiting steps.

Interestingly enough, one may also use the results obtained above for
$\tilde{S}(q,s)$ to interpret experiments performed with
rotary molecular motors, such as F1-ATPase \cite{Yoshida01,Boyer97}. In
this case, one assimilates the rotary motion of the probe attached to
the motor to the motion of {\em classical} spin in two dimensions (an
XY model), with unit length and $Q$ internal states, where $Q$ is an
integer, such that the spin rotates around the $z$-axis in one
direction by an angle equal to $2\pi/Q$. Such a model may be
appropriately referred to as a 'random-clock'.

In more mathematical terms, one defines the 2-dimensional vector
variable $\vec{\sigma}_T=(\cos(2\pi N(T)/Q),\sin(2\pi N(T)/Q))$, where
$N(T)$ is given by a renewal process with an arbitrary waiting-time
distribution $f(\zeta)$. Such a vector rotates in the 2-dimensional circle
by an angle equal to $2\pi/Q$, with the waiting-time distribution of
the renewal process $f(\zeta)$. After $Q$ rotations, the spin returns to
its original state.  It can be easily seen that the spin-spin
correlation function of such a model is given by $\langle
\vec{\sigma}_{t_1}\cdot\vec{\sigma}_{t_2} \rangle= Re
\,S(q=2\pi/Q,t_1,t_2)$, where $S(q,t_1,t_2)$ is the density-density
correlator defined above for the renewal process with waiting-time
distribution $f(\zeta)$ and $d=1$, since $q$ is
here an angular variable. In the long-time limit, $t_1,t_2\rightarrow
\infty$, where this function is solely dependent on $t=t_1-t_2$, one
can read the Laplace transform of such a quantity from equation
(\ref{GDeq4}). This transform is given by
\begin{equation}
\label{GDeq7}
{\cal L}\,\langle \vec{\sigma}_{t_1}\cdot\vec{\sigma}_{t_2}\rangle =
Re\,\left[\frac{1}{s}\,\left(1+\frac{(e^{-2\pi i/Q}-1)(1-\tilde{f}(s))}{
\langle\tau\rangle\,s\,(1-e^{-2\pi i/Q}\tilde{f}(s))}\right)\right]\,.
\end{equation}
If we were to take $Q=2$, the model would become that of an Ising spin
variable and equation (\ref{GDeq7}) reduces to the result obtained by
Godreche and Luck \cite{Godreche01}. In effect, the computation of
$\tilde{S}(q,s)$, as given by (\ref{GDeq4}), can also be obtained from
their results. We will use a different method to derive such an
expression.

In the case of F1-ATPase, the symmetry of the motor molecule implies
that $Q=3$, i.e. the motor rotates by an angle of $120$ degrees and
requires three such rotations to complete its cycle. If one were to
assume that the waiting-time distribution for each of these $120$
degrees rotations was given by a simple Poisson process, it becomes
trivial to invert the Laplace transform given in equation
(\ref{GDeq7}) and one obtains (with $t_1>t_2$)
\begin{equation}
\label{GDeq8}
\langle \vec{\sigma}_{t_1}\cdot\vec{\sigma}_{t_2}\rangle =
\cos\left(\frac{\sqrt{3} (t_1-t_2)}{2\tau}\right)\,
\exp\left(-\frac{3(t_1-t_2)}{2\tau}\right)\,,
\end{equation}
i.e., the spin-spin correlation function, despite the irreversible
character of the underlying spin model, displays an oscillatory,
albeit overdamped, behaviour.  In fact, such behaviour is always
present if $Q>2$.  Its origin is trivial, being traceable to the
circular character of the motion. Nevertheless, the study of the
motion may provide valuable information concerning the chemical cycle.
In equation (\ref{GDeq8}), such information is encoded in the turnover
time $\tau$.

\begin{figure}[htbp]
\centerline{\epsfxsize 0.75\columnwidth \epsfbox{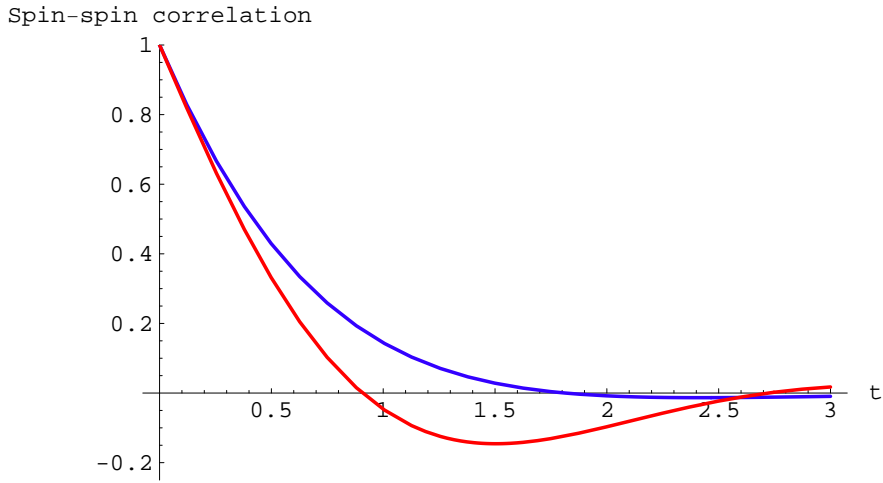}}
\caption{Spin-spin correlation functions as given by
  equations (\ref{GDeq8}) (blue curve) and (\ref{GDeq9})
  (red curve).  It is seen that the second function decays
  more slowly than the first and it also oscillates more rapidly.}
\label{F3}
\end{figure}

In reality, the $120$-degree rotation performed by F1-ATPase is
composed of two rate-limiting substeps, the first substep
corresponding to a $90$-degree rotation and the second substep to a
$30$-degree rotation  \cite{Yoshida01}. This implies that $f(\zeta)$ is
given by the convolution of two Poisson processes. In such a case, it
is still possible to invert the Laplace transform given by equation
(\ref{GDeq8}), but the resulting expression for the spin-spin
correlation function in the time domain is rather cumbersome and it is
preferable to work with its Fourier transform instead.  However, there
is one particular instance in which the spin-spin correlation function
in the time domain acquires a particularly simple form, namely when
the time-constants $\tau_1$ and $\tau_2$ are equal, i.e.
$\tau_1=\tau_2=\tau/2$.  One obtains for the spin-spin correlation
function in this case, the result (with $t_1>t_2$)
\begin{equation}
\label{GDeq9}
\langle \vec{\sigma}_{t_1}\cdot\vec{\sigma}_{t_2}\rangle =
\cos\left(\frac{\sqrt{3} (t_1-t_2)}{\tau}\right)\,
\left(\,\frac{3}{4}\,e^{-\frac{(t_1-t_2)}{\tau}}+\frac{1}{4}\,e^{-\frac{3(t_1-t_2)}{\tau}}\,\right)\,,
\end{equation}
whose form is clearly different from the one given in equation
(\ref{GDeq8}), as the period of the oscillations is now shorter than
its decay-time, i.e.  the oscillations are no longer overdamped. This
property may actually permit the experimental observation of such
oscillations. We leave the derivation of equation (\ref{GDeq9}) to
\ref{apC}. We plot below, in figure \ref{F3}, the two
functions given by (\ref{GDeq8}) and (\ref{GDeq9}) for comparison, in
terms of the variable $t=t_1-t_2$.

\begin{figure}[htbp]
  \centerline{\epsfxsize 0.75\columnwidth \epsfbox{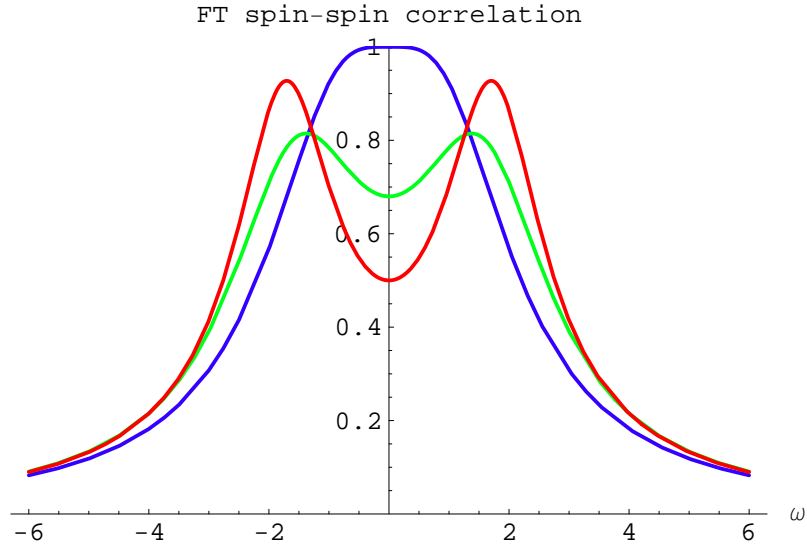}}
\caption{ Fourier transform of the spin-spin correlation
  function for a renewal process composed of two Poisson processes.
  The three cases shown correspond to the choices, $\tau_1=1$,
  $\tau_2=0$ (blue curve), $\tau_1=4/5$, $\tau_2=1/5$
  (green curve) and $\tau_1=\tau_2=1/2$ (red curve).
  Note that the area under the curve is the same for all plots. }
\label{F4}
\end{figure}

The derivation of the expression of the Fourier transform of the
spin-spin correlation function, for a general choice of the two time
constants $\tau_1$ and $\tau_2$, is also left to \ref{apC}.
Below, in figure \ref{F4}, we plot such a function for the three
distinct choices $\tau_1=1$, $\tau_2=0$, which is just the Fourier
transform of equation (\ref{GDeq8}), $\tau_1=4/5$, $\tau_2=1/5$, and
$\tau_1=\tau_2=1/2$, which is the Fourier transform of equation
(\ref{GDeq9}).  The choice of units is such that the total turnover
time $\tau=\tau_1+\tau_2=1$.  Note that the difference between
the two extreme cases $\tau_1=1$, $\tau_2=0$ and $\tau_1=\tau_2=1/2$
is more pronounced if the functions are plotted in Fourier space
(figure \ref{F4}) rather than in time-domain (figure \ref{F3}).  The
experimental situation is closer to the third case considered, i.e.
the two time-constants are approximately equal to
$1$ms \cite{Yoshida01}. Here again, the measurement of the spin-spin
correlation, followed by a Fourier transformation of the signal, is
the preferred method for the extraction of the chemical cycle's time
constants, since one can fit the experimentally measured signal with
the result of equation (\ref{apCeq3}), which then yields $\tau_1$ and
$\tau_2$. 

As we have mentioned above, the measurement of the spin-spin
correlation function should ideally be performed using a 
fluorescent dye, since such a probe does not affect 
the rotation of F1-ATPase. The visualisation
of the single fluorophore, coupled to the rotating 
$\gamma$ sub-unit of F1-ATPase, is performed by excitation 
of the fluorophore with circular
polarized light and by measuring the intensities, $H$ and $V$,
of the light emitted by the molecule 
along two perpendicular directions of polarisation 
in a dual-view apparatus \cite{Adachi00}.
The geometry of the experimental setup is such that the polarisation of the 
observed light is parallel to the dipolar moment of the fluorophore,
and is therefore determined by the instantaneous
angle of rotation $\theta(T)=\frac{2\pi}{3}\,N(T)$,
of the $\gamma$ sub-unit with respect to the
fixed $\alpha\beta$ sub-units \cite{Adachi00}, 
where we take $N(T)$ to be given by a renewal process, as above.
Hence, one concludes that $H(T)\propto \cos^2(\frac{2\pi}{3}N(T)+\phi)$, 
$V(T)\propto \sin^2(\frac{2\pi}{3}N(T)+\phi)$, where $\phi$
is the angle between the horizontal polarisation direction and
the nearest polarisation direction along which light is emitted,
with $\phi=18^{o}$ in the experiments of Adachi and 
co-workers \cite{Adachi00}. 

With such definitions in hand,
we define $h(t_2)=\pm\sqrt{\frac{H(t_2)}{H(t_2)+V(t_2)}}$, 
$v(t_2)=\pm\sqrt{\frac{V(t_2)}{H(t_2)+V(t_2)}}$,
the sign in these definitions being determined by the value of the
instantaneous polarisation $P(t_2)=h^2(t_2)-v^2(t_2)$ \cite{pol_signs}.
Using the same definition for
$h(t_1)$, $v(t_1)$ at a later time $t_1$, 
one concludes that 
\begin{equation}
\label{GDeq9a}
\!\!\!\!
\langle \vec{\sigma}_{t_1}\cdot\vec{\sigma}_{t_2}\rangle=
\langle \cos [2\pi(N(t_1)-N(t_2))/3]\rangle=
\langle h(t_1)h(t_2)\rangle+\langle v(t_1)v(t_2)\rangle,
\end{equation}
an equality that shows that the spin-spin correlation function can be 
related to the average value of a directly measurable quantity. 
One should
note that the experiments of Adachi et al. \cite{Adachi00} were
performed at low ATP concentrations where ATP binding is a rate-limiting
step, with typical binding times of the order of seconds. If one
wishes to measure the characteristic times $\tau_1$ and $\tau_2$,
one needs to perform the experiment at much higher ATP concentrations
and with a time-resolution which is three orders of magnitude higher
than in the experiment of Adachi and co-workers ($\tau_1\approx
\tau_2\approx 1$ ms). Achieving such a
resolution may constitute an insurmountable practical problem, even 
if the method theoretically permits an unlimited time-resolution
 \cite{Thanks_2_referee}.

\section{Definition of the renewal process and associated quantities}
\label{secIII}
A renewal process is defined by a random variable $N(u)$ that is
incremented by one unit at random times $u=T_1< T_2<\dots <T_{\cal N}$
where, in the simplest case and to which we will attain, the length of
the intervals $T_1, T_2-T_1,\dots, T_{\cal N}-T_{{\cal N}-1}$ is taken
from a given waiting time distribution $f(\zeta)$ such that $f(\zeta)$ is
positive and $\int_{0}^{\infty}\,d\zeta\,f(\zeta)=1$ (one assumes that the
process starts at $u=0$ with $N(0)=0$).  Mathematically, this can be
written as
\begin{equation}
N(u)=n\;\;\mbox{if}\;\;T_n\leq u<T_{n+1}\,,
\label{eq1}
\end{equation}
with $T_0=0$. If one defines $F_k(T)$ to be the probability that at
least $k$ increments have occurred up to time $T$ irrespective of
whether more increments have occurred or not, then it is easy to
show \cite{Cox} that $F_k(T)$ is given by
\begin{eqnarray}
F_0(T)&=&1\\
\label{eq2}
F_k(T)&=&\int_{0}^{T}\,d\zeta\,F_{k-1}(T-\zeta)\,f(\zeta)\;\;k>0\,.
\label{eq3}
\end{eqnarray}
One may equally regard the renewal process as a totally asymetric
random-walk of a structureless particle, in which case one speaks of
the probability of observing at least $k$ forward steps of such a
particle up to time $T$.  As discussed above, this particle is
supposed to represent a molecular motor, with the waiting
time-distribution $f(\zeta)$ being determined by the internal chemical
cycle of the motor.

In particular, it follows from this definition that
$F_1(T)=\int_{0}^{T}\,d\zeta\,f(\zeta)$ is the cumulative probability to
observe an increment of $N(u)$ in the time interval $[0,T[$. The
probability that exactly $k$ increments have occurred up to time $T$
is then given by
\begin{equation}
P_k(T)=F_k(T)-F_{k+1}(T)\,.
\label{eq4}
\end{equation}
The average number of increments of $N(u)$ (or the average number of
steps of the particle) up to time $T$, which we designate by $m(T)$,
is given by $m(T) =\langle
N(T)\rangle=\sum_{k=0}^{\infty}\,k\,P_k(T)$. Using equation
(\ref{eq4}) for $P_k(T)$ and expressing $F_{k+1}(T)$ in terms of
$F_k(u)$ at earlier times through equation (\ref{eq3}), it follows
that $m(T)$ necessarily obeys the integral equation
\begin{equation}
m(T)=F_1(T)+\int_{0}^{T}\,d\zeta\,m(T-\zeta)\,f(\zeta)\,,
\label{eq5}
\end{equation}
which is known in the literature as the 'renewal equation', $m(T)$
being called the 'renewal function'. For reasons that will become
clear below, we prefer to work with the derivative of $m(T)$ and we
write $l(T)=m'(T)$, which we will also call the renewal function.
Differentiating the above equation with respect to $T$ and given that
$m(0)=0$, one obtains
\begin{equation}
l(T)=f(T)+ \int_{0}^{T}\,d\zeta\,l(T-\zeta)\,f(\zeta)\,,
\label{eq6}
\end{equation}
i.e. the equation has the same form as (\ref{eq5}), but the
non-homogeneous term is now given by $f(T)$ rather than by $F_1(T)$. From 
the above equation, it also follows that $l(0)=f(0)$. If one
considers a Poisson process, where $f(\zeta)=e^{-\zeta/\tau}/\tau$, it is
trivial to verify that $l(T)=1/\tau$ is the solution of equation
(\ref{eq6}) and one has $m(T)=T/\tau$, i.e.  the average number of
steps of the walker increases linearly with time. For more general
forms of the waiting-time distribution, the renewal equation can still
be solved by Laplace transformation, and one obtains
\begin{equation}
\tilde{l}(s)=\frac{\tilde{f}(s)}{1-\tilde{f}(s)}\,, 
\label{eq6A}
\end{equation}
where $\tilde{l}(s)$ and $\tilde{f}(s)$ are, respectively, the Laplace
transforms of $l(T)$ and of $f(T)$.  In order to obtain $l(T)$ from
this expression, one needs to invert the Laplace transform, an
operation one can only perform in a limited number of cases.
Nevertheless, Blackwell's renewal theorem \cite{Cox} asserts that
provided that $\langle \tau\rangle=\int_0^\infty\,d\zeta\,\zeta\,f(\zeta)$ is
finite, then $l(T)=\frac{1}{\langle\tau\rangle}$ asymptotically at
large time $T$.

Having defined the random process that will be the object of our
study, we now proceed to define the quantities we wish to compute. We
will first consider the restriction of the renewal process given above
such that exactly ${\cal N}$ increments have occurred between time
$u=0$ and time $u=T$, i.e. the times of the increments are such that
$0<T_1<T_2<\dots<T_{\cal N}<T<T_{{\cal N}+1}$ and we will consider
correlation functions of the random variable \cite{Honerkamp}
\begin{equation}
{\cal J}(u)=\sum_{i=1}^{\cal N}\,g(u-T_i)\,,
\label{eq7}
\end{equation}
where $0\leq u<T$ and where $g(u)$ is a function with at most a finite
number of discontinuities. If ${\cal N}=0$, we define ${\cal
  J}(u)\equiv 0$ for all $0\leq u<T$.  Note that
$N(u)=\sum_{i=1}^{\cal N}\,\theta(u-T_i)$, where $\theta(u)=0$ if
$u<0$, $\theta(u)=1$ if $u\geq 0$, is of this form, and therefore the
number of increments $N(u)$ (or the number of steps which the particle
has given) corresponds to a particular case of (\ref{eq7}) in which
one takes $g(u)=\theta(u)$.  Thus, we wish to determine the value of
the correlation functions $\langle {\cal J}(t_1)\,{\cal
  J}(t_2)\,\dots\,{\cal J}(t_m)\rangle_{\cal N}$, where $m\geq 1$ and
where the averaging is over the distribution of the times $T_1$ to
$T_{\cal N}$. For simplicity, we will restrict ourselves to $m=1$ and
$m=2$ in the particular examples discussed, though the formalism is
valid for arbitrary $m$.

If one defines the probability-generating functional $Z_{\cal N}[j]$ as
\begin{equation}
Z_{\cal N}[j]=\left\langle \exp\left(i\int_{0}^{T}\,dx\,j(x)\,{\cal
    J}(x)\right)\right\rangle_{\cal N}\,,
\label{eq8}
\end{equation}
where the averaging is performed over the increment times
$T_1,\dots,T_{\cal N}$, then it follows that the computation of
$Z_{\cal N}[j]$ is equivalent to the computation of all correlation
functions, which can be obtained by functionally differentiating
$Z_{\cal N}[j]$ with respect to $j$. Note that we choose the
normalisation of $Z_{\cal N}[j]$ to be given by
\begin{equation}
Z_{\cal N}[j=0]=P_{\cal N}(T)\,,
\label{eq9}
\end{equation}
where $P_{\cal N}(T)$ is the probability of occurrence of exactly
${\cal N}$ increments up to time $T$, introduced above.

One can now generalise the concept of probability-generating functional
to situations in which the total number of increments is not
fixed by introducing the (grand-canonical) functional \cite{Honerkamp}
\begin{equation}
{\cal Z}[j]=\sum_{{\cal N}=0}^{\infty}\,Z_{\cal
  N}[j]\,.
\label{eq10}
\end{equation}
The functional ${\cal Z}[j]$ can again be functionally differentiated
with respect to $j$ in order to obtain the relevant correlation
function.  Note that if one takes $j=0$, one obtains
\begin{equation}
{\cal Z}[j=0]=\sum_{{\cal N}=0}^{\infty}\,P_{\cal N}(T)=1\,,
\label{eq11}
\end{equation}
i.e. the probability-generating functional is equal to one due 
to normalisation of
the total probability. This result justifies the normalisation chosen
above for $Z_{\cal N}[j]$. In the next section, we will develop a
technique which will allow us to compute the correlators of ${\cal
  J}(u)$ order by order.

\section{Integral equation for the probability-generating functional and
  associated multiple-time correlation functions}
\label{secIV}

We first consider the case in which exactly ${\cal N}>0$ increments
have occurred up to time $T$. The exclusive probability density (also
known as Janoussi local density) \cite{Macchi71,Daley_Vere-Jones} for
the increments to occur at times $T_1<T_2<\dots<T_{\cal N}<T$ is given
by
\begin{eqnarray}
p_{\cal N}(T_1,T_2,\dots,T_{\cal N})&=&P_{0}(T-T_{\cal N})f(T_{\cal
  N}-T_{{\cal N}-1})\dots f(T_2-T_1)f(T_1)\,,
\label{eq12}
\end{eqnarray}
where $P_{0}(T-T_{\cal N})=1-\int_{0}^{T-T_{\cal N}}\,d\zeta\,f(\zeta)$ is the
probability that no increment has ocurred between $T_{\cal N}$ and
$T$, as given by equation (\ref{eq4}) (with $k=0$). The probability-generating
functional $Z_{\cal N}[j]$ is given in terms of this quantity by
\begin{eqnarray}
\!\!\!\!\!\!\!\!\!\!\!\!
Z_{\cal N}[j]&=&\int_{0}^{T}\,dT_{\cal N}\,\int_{0}^{T_{\cal N}}\,dT_{{\cal
    N}-1}\,\dots\,\int_{0}^{T_2}\,dT_{1}\,p_{\cal N}(T_1,T_2,\dots,T_{\cal
    N})\,e^{i\int_{0}^{T}\,dx\,j(x)\,{\cal J}(x)}\nonumber\\
&=&\int_{0}^{T}\,dT_{\cal N}\,P_{0}(T-T_{\cal
    N})\,e^{i\int_{0}^{T}\,dx\,j(x)\,g(x-T_{\cal N})}\nonumber\\
&&\int_{0}^{T_{\cal N}}\,\,dT_{{\cal
    N}-1}\,f(T_{\cal
  N}-T_{{\cal N}-1})\,e^{i\int_{0}^{T}\,dx\,j(x)\,g(x-T_{{\cal N}-1})}\,
\ldots\nonumber\\
&&\int_{0}^{T_2}\,dT_{1}\,f(T_2-T_1)\,e^{i\int_{0}^{T}\,dx\,j(x)\,g(x-T_{1})}\,f(T_1)\,,
\label{eq13}
\end{eqnarray}
where we have used the identity $e^{i\int_{0}^{T}\,dx\,j(x)\,{\cal
    J}(x)}=\prod_{l=1}^N\,e^{i\int_{0}^{T}\,dx\,j(x)\,g(x-T_{l})}$.
Note that if $j(x)=0$, the functional $Z_{\cal N}[j=0]$ reduces to
$P_{\cal N}(T)$, as stated above. This equation can be written in a
more condensed form if one introduces the following set of
functionals, defined recursively by
\begin{eqnarray}
l_0(j,u)&=&f(u)\nonumber\\
l_{\cal N}(j,u)&=&\int_{0}^{u}\,\,dv
\,f(u-v)\,e^{i\int_{0}^{T}\,dx\,j(x)\,g(x-v)}\,l_{{\cal N}-1}(j,v)\,.
\label{eq14}
\end{eqnarray}
Using this notation, it is then easy to see that one can write the
functionals $Z_{\cal N}[j]$ in the form
\begin{eqnarray}
Z_0[j]&=&P_0(T)\nonumber\\
Z_{\cal N}[j]&=&\int_{0}^{T}\,\,du
\,P_0(T-u)\,e^{i\int_{0}^{T}\,dx\,j(x)\,g(x-u)}\,l_{{\cal
    N}-1}(j,u)\;\; 
\mbox{if}\;\;{\cal N}>0\,.
\label{eq15}
\end{eqnarray}
Since we wish to compute the functional ${\cal Z}[j]$, we need to
performed the summation defined in equation (\ref{eq10}). If one now
introduces the functional
\begin{equation}
l(j,u)=\sum_{{\cal N}=0}^{\infty}\,l_{\cal
  N}(j,u)\,,
\label{eq16}
\end{equation}
then one can show from equation (\ref{eq14}) that $l(j,u)$ obeys the
integral equation
\begin{eqnarray}
l(j,u)&=&f(u)+\int_{0}^{u}\,\,dv
\,f(u-v)\,e^{i\int_{0}^{T}\,dx\,j(x)\,g(x-v)}\,l(j,v)\,.
\label{eq17}
\end{eqnarray}
Now, using the recursion relation (\ref{eq15}), it is easy to show
that ${\cal Z}[j]$ can be expressed in terms of $l(j,u)$ by
\begin{eqnarray}
{\cal Z}[j]&=&P_0(T)+\int_{0}^{T}\,\,du
\,P_0(T-u)\,e^{i\int_{0}^{T}\,dx\,j(x)\,g(x-u)}\,l(j,u)\,.
\label{eq18}
\end{eqnarray}
As they stand, equations (\ref{eq17}) and (\ref{eq18}) are of little
use, since it is usually not possible to solve (\ref{eq17}) for
$l(j,u)$. However, if one takes $j(x)=0$ in equation (\ref{eq17}), one
obtains, after changing the integration variable from $v\rightarrow
u-v$, the equation
\begin{eqnarray}
l(0,u)&=&f(u)+\int_{0}^{u}\,\,dv
\,l(0,u-v)\,f(v)\,,
\label{eq19}
\end{eqnarray}
with the initial condition $l(0,0)=f(0)$. But this is precisely
equation (\ref{eq6}) for the renewal function $l(u)$, with the same
initial condition. Therefore, we conclude that $l(0,u)=l(u)$, a result
which justifies the notation we are using. Now, we express $f(u)$ and
$f(u-v)$ in equation (\ref{eq17}) in terms of $l(u)$, using
(\ref{eq19}).  After interchanging the limits of integration in the
resulting double integral and using (\ref{eq17}), one obtains the
integral equation
\begin{eqnarray}
l(j,u)&=&l(u)+\int_{0}^{u}\,\,dv
\,l(u-v)\,\left(\,e^{i\int_{0}^{T}\,dx\,j(x)\,g(x-v)}-1\right)\,l(j,v)\,,
\label{eq20}
\end{eqnarray}
which shows explicitly that $l(j,u)$ reduces to $l(u)$ when $j(x)=0$.
Now, substituting $P_0(T)=1-\int_0^T\,du\,f(u)$,
$P_0(T-u)=1-\int_u^T\,dv\,f(v-u)$ in equation (\ref{eq18}) and
interchanging the limits of integration in the resulting double
integral, one obtains, after using equation (\ref{eq17})
\begin{eqnarray}
{\cal Z}[j]&=&1+\int_{0}^{T}\,\,du
\left(\,e^{i\int_{0}^{T}\,dx\,j(x)\,g(x-u)}-1\right)\,l(j,u)\,,
\label{eq21}
\end{eqnarray}
which shows explicitly that ${\cal Z}[j=0]=1$. Equations (\ref{eq20})
and (\ref{eq21}) are the main results of this section.  In fact, these
two equations are equivalent to the expansion of the probability-generating
functional ${\cal Z}[j]$ in terms of the conditional probability
densities for a renewal process, see \cite{Daley_Vere-Jones} for
details. The above form is more convenient for explicit calculations,
as shown below.

Before we discuss the general solution of (\ref{eq20}), let us apply
our results to compute the functional ${\cal Z}[j]$ for the particular
case of the simple Poisson process. Given that $l(u)=1/\tau$, one concludes,
substituting this result in equation (\ref{eq20}), that $l(j,u)$ obeys
the integral equation
\begin{eqnarray}
l(j,u)&=&\frac{1}{\tau}+\int_{0}^{u}\,\,\frac{dv}{\tau}
\,\left(\,e^{i\int_{0}^{T}\,dx\,j(x)\,g(x-v)}-1\right)\,l(j,v)\,.
\label{eq22}
\end{eqnarray}
Differentiating this equation with respect to $u$, one obtains the
first-order linear differential equation
\begin{eqnarray}
\frac{d l(j,u)}{du}&=&\frac{1}{\tau}
\,\left(\,e^{i\int_{0}^{T}\,dx\,j(x)\,g(x-u)}-1\right)\,l(j,u)\,.
\label{eq23}
\end{eqnarray}
with the initial condition $l(j,0)=1/\tau$. Such 
an equation can be quickly integrated
by separation of variables, yielding the solution
\begin{eqnarray}
l(j,u)&=&\frac{1}{\tau}\exp\left(\int_{0}^{u}\,\,\frac{dv}{\tau}
\,\left(\,e^{i\int_{0}^{T}\,dx\,j(x)\,g(x-v)}-1\right)\,\right)\,.
\label{eq24}
\end{eqnarray}
This solution, when substituted in equation (\ref{eq21}), yields
\begin{eqnarray}
{\cal Z}[j]&=&\exp\left(\int_{0}^{T}\,\,\frac{du}{\tau}
\,\left(\,e^{i\int_{0}^{T}\,dx\,j(x)\,g(x-u)}-1\right)\,\right)\,,
\label{eq25}
\end{eqnarray}
which is a well known result, being given in, e.g. \cite{Honerkamp}.
However, this is in fact the only case for which a closed expression
for ${\cal Z}[j]$ can be found, as one can no longer reduce the
integral equation (\ref{eq20}) to a first-order differential equation
if one chooses a different waiting-time distribution. Neither can one
solve (\ref{eq20}) using a Laplace transform because the source term
$e^{i\int_{0}^{T}\,dx\,j(x)\,g(x-v)}$ in (\ref{eq20}) is not 
time-translation invariant. Such difficulties stem from the fact
that equation (\ref{eq20}) is a Volterra integral equation for which
no analytical solution is known in the general case.

Nevertheless, one can show that all correlation functions can be
obtained from equations (\ref{eq20}) and (\ref{eq21}) in a closed
form, provided one knows the renewal function $l(u)$. This point can
be made clearer through the discussion of two examples which we will
use extensively in our applications of renewal processes to the study
of the dynamics of molecular motors. We wish to compute the average
value $\langle {\cal J}(t_1)\rangle$ and the two-point correlation
function $\langle {\cal J}(t_1)\,{\cal J}(t_2)\rangle$. In the first
case, one has
\begin{eqnarray}
\langle {\cal J}(t_1)\rangle &=&-i\left.\frac{\delta {\cal Z}[j]}{\delta j(t_1)}\right|_{j=0}
\nonumber\\
&=&\left(\,\int_{0}^{T}\,du\,g(t_1-u)\,e^{i\int_{0}^{T}\,dx\,j(x)\,g(x-u)}\,l(j,u)\right.\nonumber\\
&&\mbox{}\left.-i\int_{0}^{T}\,du\,\left(\,e^{i\int_{0}^{T}\,dx\,j(x)\,g(x-u)}-1\right)\,
\frac{\delta l(j,u)}{\delta j(t_1)}\,\right)_{j=0}\nonumber\\
&=&\int_{0}^{T}\,du\,g(t_1-u)\,l(u)\,,
\label{eq26}
\end{eqnarray}
since the second term is zero if $j(x)=0$ and $l(j=0,u)=l(u)$.
Therefore, the average value $\langle {\cal J}(t_1)\rangle$ can be
expressed as an integral involving the known functions $g(u)$ and
$l(u)$. In particular, if ${\cal J}(t_1)=N(t_1)$, which is the number
of increments which has occurred up to time $t_1$ or the displacement
of the particle up to time $t_1$, then one has $g(u)=\theta(u)$ and
one obtains
\begin{equation}
\langle N(t_1)\rangle =\int_{0}^{t_1}\,du\,l(u)\;=\;m(t_1)\,,
\label{eq27}
\end{equation}
which is just the definition of the renewal function $m(t_1)$ given in
the previous section.

The procedure used to compute the two point correlator $\langle {\cal
  J}(t_1)\,{\cal J}(t_2)\rangle$ is analogous, but it also involves
the functional derivative $\frac{\delta l(j,u)}{\delta j(t_{1,2})}$.
One has
\begin{eqnarray}
\!\!\!\!\!\!\!\!\!\!\!\!\!\!\!\!\!\!\!\!\!
\langle {\cal J}(t_1)\,{\cal J}(t_2)\rangle \!&=&\!
-\left.\frac{\delta^2 {\cal Z}[j]}{\delta j(t_1)\delta j(t_2)}\right|_{j=0}
=\int_{0}^{T}du\,g(t_1-u)\,g(t_2-u)\,l(u)\nonumber\\
&&\mbox{}-i\int_{0}^{T}du\,g(t_1-u)\left.\frac{\delta l(j,u)}{\delta
  j(t_2)}\right|_{j=0}-i\int_{0}^{T}du\,g(t_2-u)\left.
\frac{\delta l(j,u)}{\delta j(t_1)}\right|_{j=0}\!\!\!\!\!\!,
\label{eq28}
\end{eqnarray}
since the term which involves the functional derivative
$\frac{\delta^2 l(j,u)}{\delta j(t_1)\delta j(t_2)}$ is zero if
$j(x)=0$. Now, in order to compute the functional derivative
$\frac{\delta l(j,u)}{\delta j(t_{1})}$ at $j(x)=0$, one functionally
differentiates equation (\ref{eq20}) with respect to $j(t_{1})$ and
sets $j(x)=0$. One obtains
\begin{eqnarray}
\!\!\!\!\!\!\!\!\!\!\!\!
\left.\frac{\delta l(j,u)}{\delta j(t_1)}\right|_{j=0}
&=&i\left(\,\int_{0}^{u}\,dv\,l(u-v)\,\,g(t_1-v)\,e^{i\int_{0}^{T}\,dx\,j(x)\,g(x-v)}\,l(j,v)\right.\nonumber\\
&&\mbox{}\left.+\int_{0}^{u}\,dv\,l(u-v)\,\left(\,e^{i\int_{0}^{T}\,dx\,j(x)\,g(x-v)}-1\right)\,
\frac{\delta l(j,v)}{\delta j(t_1)}\,\right)_{j=0}\nonumber\\
&=&i\,\int_{0}^{t}\,dv\,l(u-v)\,g(t_1-v)\,l(v)\,,
\label{eq29}
\end{eqnarray}
since the second term is again zero if $j(x)=0$ and $l(j=0,v)=l(v)$.
The expression for the derivative $\frac{\delta l(j,u)}{\delta j(t_{2})}$
is simply obtained from equation (\ref{eq29}) by interchanging $t_1$ by $t_2$. 

It should be clear from equation (\ref{eq26}) that in order to compute
the $m$-th variational derivative of ${\cal Z}[j]$ at $j(x)=0$, one
needs only the derivative of order $m-1$ of $l(j,u)$, because the term
which involves the $m$-th derivative of $l(j,u)$ is zero in that case,
since $e^{i\int_{0}^{T}\,dx\,j(x)\,g(x-u)}-1=0$ if $j(x)=0$.  More
importantly, it should also follow from the structure of equation
(\ref{eq29}) that in order to compute the $m$-th variational
derivative of $l(j,u)$ at $j(x)=0$, one only needs the derivative of
order $m-1$ of $l(j,u)$, because the term which involves the $m$-th
derivative of $l(j,u)$ is zero for the same reason. This implies that
one can express such variational derivatives in terms of quantities
which were previously calculated, assuming that we know the function
$l(u)$. Thus, the solution of equation (\ref{eq6}) is sufficient to
compute all correlation functions in a recursive manner,
see \cite{Daley_Vere-Jones} for a rigorous proof.

Finally, one can obtain a closed expression for $\langle {\cal
  J}(t_1)\,{\cal J}(t_2)\rangle$ by substituting the result for
$\left.\frac{\delta l(j,u)}{\delta j(t_{1,2})}\right|_{j=0}$ given
in equation (\ref{eq29}) in equation (\ref{eq28}). One has that
$\langle {\cal J}(t_1)\,{\cal J}(t_2)\rangle$ is given by
\begin{eqnarray}
\langle {\cal J}(t_1)\,{\cal J}(t_2)\rangle &=&\int_{0}^{T}\,du\,g(t_1-u)\,g(t_2-u)\,l(u)\nonumber\\
&&\mbox{}+\int_{0}^{T}\,du\,g(t_1-u)\,\int_{0}^{u}\,dv\,l(u-v)\,g(t_2-v)\,l(v)\nonumber\\
&&\mbox{}+\int_{0}^{T}\,du\,g(t_2-u)\,\int_{0}^{u}\,dv\,l(u-v)\,g(t_1-v)\,l(v)\,.
\label{eq30}
\end{eqnarray}
If ${\cal J}(t_1)=N(t_1)$, ${\cal J}(t_2)=N(t_2)$, then $g(u)=\theta(u)$ and
we obtain for the correlation function $\langle N(t_1)\,N(t_2)\rangle$, the result
\begin{eqnarray}
\langle N(t_1)\,N(t_2)\rangle &=&\int_{0}^{\mbox{\scriptsize
    min}(t_1,t_2)}\,du\,l(u)
+\int_{0}^{t_1}\,du\,\int_{0}^{\mbox{\scriptsize min}(u,t_2)}\,dv\,l(u-v)\,l(v)\nonumber\\
&&\mbox{}+\int_{0}^{t_2}\,du\,\int_{0}^{\mbox{\scriptsize min}(u,t_1)}\,dv\,l(u-v)\,l(v)\,.
\label{eq31}
\end{eqnarray}
In the context of the study of molecular motors, it is more
appropriate to consider the mean-square deviation of the displacement
of the particle between time $t_2$ and $t_1$ ($t_1>t_2$), i.e. the
correlation function $\langle
(N(t_1)-N(t_2))^2\rangle_{\mbox{\scriptsize conn}}= \langle
(N(t_1)-N(t_2))^2\rangle -\langle N(t_1)-N(t_2)\rangle^2$, as
explained in section \ref{secII}. If we use equations (\ref{eq27}) and
(\ref{eq31}), we obtain, after some trivial manipulations,
\begin{equation}
\!\!\!\!\!\!\!\!\!\!\!\!\!\!\!
\langle (N(t_1)-N(t_2))^2\rangle_{\mbox{\scriptsize conn}}=
\int_{t_2}^{t_1}du\,l(u)
+2\!\int_{t_2}^{t_1}du\,l(u)\int_{u}^{t_1}dv\,(l(v-u)-l(v)).
\label{eq32}
\end{equation}
In the case of a Poisson process, $l(u)=1/\tau$ and this expression
reduces to the first term, i.e. we obtain $\langle
(N(t_1)-N(t_2))^2\rangle_{\mbox{\scriptsize conn}}=(t_1-t_2)/\tau$,
which is a well-known result for the simple Poisson process. The 
expression for the full functional obtained above in equation
(\ref{eq25}) allows one to compute correlation functions of any
order, but only in this simple case.

A particular important limit of (\ref{eq32}) is the one in which
$t_2,t_1\rightarrow \infty$, but in which the time difference
$t=t_1-t_2$ is kept finite, since ${\cal
  C}(t)=\lim_{t_1,t_2\rightarrow\infty} \langle
(N(t_1)-N(t_2))^2\rangle_{\mbox{\scriptsize conn}}$ becomes a function
of $t$ only, i.e. one recovers a form of time translation invariance.
In this limit, $l(u)\rightarrow \frac{1}{\ave{\tau}}$ and one obtains,
substituting this result above, the following result for  ${\cal C}(t)$ 
\begin{eqnarray}
{\cal C}(t)&=&\frac{t}{\ave{\tau}}+\frac{2}{\ave{\tau}}\,
\int_{0}^{t}\,du\,(t-u)\,\left(l(u)-\frac{1}{\ave{\tau}}\right)\,.
\label{eq33}
\end{eqnarray}
If $t\ll \ave{\tau}$, it is easy to see that the second term of this
equation is at least $O(t^2)$ and therefore ${\cal C}(t) \approx
\frac{t}{\ave{\tau}}$, if $t\ll \ave{\tau}$.  This observation
justifies the first line of equation (\ref{GDeq3}).  In order to
extract the long-time behaviour of ${\cal C}(t)$, one needs to
consider instead the Laplace transform $\tilde{{\cal C}}(s)$ of ${\cal
  C}(t)$. One has, from equation (\ref{eq33}),
\begin{eqnarray}
\tilde{{\cal C}}(s)&=&
\frac{1}{\ave{\tau}s^2}\,\left(\,\frac{1+\tilde{f}(s)}{1-\tilde{f}(s)}-\frac{2}{\ave{\tau}s}\,\right)\,,
\label{eq34}
\end{eqnarray}
where we have used equation (\ref{eq6A}) to express $\tilde{l}(s)$ in
terms of $\tilde{f}(s)$. If $\tilde{f}(s)$ is analytic a $s=0$, i.e.
if all moments of the distribution $f(\zeta)$ exist, one can write, for
small $s$
$\tilde{f}(s)=1-\ave{\tau}\,s+\frac{\ave{\tau^2}}{2!}\,s^2-\frac{\ave{\tau^3}}{3!}\,s^3+\dots$.
Performing a Taylor expansion of (\ref{eq34}) at $s=0$ using such a
result for $\tilde{f}(s)$ and keeping only the most divergent terms,
one obtains
\begin{eqnarray}
\tilde{{\cal C}}(s)&\approx& \frac{r}{\ave{\tau}s^2}+\frac{C}{s}\,,
\label{eq35}
\end{eqnarray}
with $r=\frac{\ave{\tau^2}-\ave{\tau}^2}{\ave{\tau}^2}$ and
$C=\frac{\ave{\tau^2}^2}{2\ave{\tau}^4}-
\frac{\ave{\tau^3}}{3\ave{\tau}^3}$, where
$\ave{\tau^2}=\int_0^\infty\,d\zeta \,\zeta^2 f(\zeta)$ and
$\ave{\tau^3}=\int_0^\infty\,d\zeta\,\zeta^3 f(\zeta)$.  
Furthermore, one can write $C$ in the form given by 
equation (\ref{GDeq3A}), using the definitions of the
connected second- and third-moment $r$ and $\delta$, given above.
The behaviour of ${\cal C}(t)$ for $t\gg \ave{\tau}$ is 
determined by the behaviour of the
Laplace transform $\tilde{{\cal C}}(s)$ at small $s$, as given by
(\ref{eq35}) and one can directly identify the coefficients of the
most divergent terms of ${\cal C}(t)$ from those in equation
(\ref{eq35}). One thus concludes that ${\cal C}(t)$ is given, in this
limit, by the second line of equation (\ref{GDeq3}).

Let us now consider the case in which the waiting-time distribution of
the renewal process is given by a convolution of ${\cal M}$ Poisson
processes, each of which is supposed to represent a rate-limiting step
of the molecular motor.  Mathematically, this is expressed by saying
that such function, which we denote by $f_{{\cal M}}(\zeta)$, is given
recursively by
\begin{equation}
f_{\cal M}(\zeta)=\int_{0}^{\zeta}\,\frac{d\eta}{\tau_{\cal M}}\,e^{-\frac{\zeta-\eta}{\tau_{\cal
    M}}}\,f_{{\cal M}-1}(\eta) \;\;\mbox{if}\;\; {\cal M}>1\,,
\label{eq36}
\end{equation}
where $f_{{\cal M}-1}(\eta)$ is the waiting-time distribution of a process with
${\cal M}-1$ rate-limiting steps and $f_1(\zeta)=e^{-\zeta/\tau_{1}}/\tau_1$.
Given that the Laplace transform of a single Poisson process is given
by $\tilde{f}_1(s)=(1+\tau_1s)^{-1}$, one concludes from 
equation (\ref{eq36}) and the convolution theorem that
the Laplace transform of $f_{\cal M}(\zeta)$ is given by
\begin{equation}
\tilde{f}_{\cal M}(s)=\prod_{k=1}^{\cal M}\,\frac{1}{1+s\tau_k}\,.
\label{eq37}
\end{equation}
Since $\ave{\tau}=-\tilde{f}'(0)$,
$\ave{\tau^2}=\tilde{f}''(0)$ and $\ave{\tau^3}=-\tilde{f}^{(3)}(0)$,
one concludes, differentiating equation (\ref{eq37}), that
$\ave{\tau}$, $r$ and $C$ are given, respectively, by equations
(\ref{GDeq0}), (\ref{GDeq2}) and (\ref{GDeq3A})
with $\delta=\frac{\sum_{i=1}^{\cal M}\tau_i^3}{\left(\sum_{i=1}^{\cal M}
\tau_i\right)^3}$. The form which
$\tilde{f}_{\cal M}(s)$ takes for ${\cal M}=2,3$ even allows one to
invert the Laplace transform $\tilde{l}(s)$ and obtain $l(u)$ in these
two cases. One can therefore determine ${\cal C}(t)$ for arbitrary
time $t$, using equation (\ref{eq33}).  This calculation is explicitly
performed in \ref{apA}, as stated above.

In the above derivation, we have assumed that all
moments of the distribution $f(\zeta)$ exist, in other
words that the Laplace transform $\tilde{f}(s)$ is an
analytic function at $s=0$. Such assumption is enterily 
justified when discussing the motion of a molecular motor 
in a homogeneous environment, as is the case of myosin V or kinesin.
On the other hand, if one considers the motion of tracer
particles in a rapidly rotating fluid \cite{Weeks96},
one is led to consider a renewal process with a 
waiting-time distribution with fat tails, which reflects 
the diverging sticking-times of the tracer particles. 
In a biological context, some of these results may also 
be applicable to the study of the 
motion of RNA-polymerase along a DNA-strand \cite{Kafri04}.

It is beyond our means to provide a complete discussion of
this issue, However, one may consider a very 
simple model for a waiting-time distribution with fat tails,
namely we take $f(\zeta)=\frac{(\nu-1)A^{\nu-1}}{(A+\zeta)^\nu}$, where
$\nu>1$, such that $f(\zeta)$ is normalisable. For $2<\nu<3$, the
assymptotic decay of $f(\zeta)$ is the same as that of a stable 
L\'evy distribution \cite{Bouchaud_TRF}. The Laplace transform 
of $f(\zeta)$ can be readily calculated, at least in the limit 
of small $s$, and one may carry through a significant number
of calculations with this simple model.

One should start by considering the consequences of taking
$\nu\leq 2$, in which case the distribution $f(\zeta)$ has an infinite 
first moment $\ave{\tau}$. In such a case, one can neither 
obtain equation (\ref{eq33})
from equation (\ref{eq32}) nor equation (\ref{GDeq4}) from 
equation (\ref{eq44}) (see below),
since we are  assuming  that $\ave{\tau}$ is finite in the derivation
of these results.
Likewise, the equations that follow from equation (\ref{eq33}), such as 
equation (\ref{eq34}), and from equation (\ref{GDeq4}), such as 
equation (\ref{GDeq5}),
are equally invalid. In physical terms, the assumption that fails 
is that one can consider the existence of a time-translation invariant regime, 
$t_1,t_2\gg \ave{\tau}$, for which the correlation functions
$\langle (N(t_1)-N(t_2))^2\rangle_{\mbox{\scriptsize conn}}$ and $S(q,t_1,t_2)$
depend only on $t=t_1-t_2$, since $\ave{\tau}$ is 
formally infinite. 

In the case in which $\nu>2$, a time-translational regime does arise and
our results hold through, except equation (\ref{eq35}) which relies on the
assumption that $\tilde{f}(s)$ is analytical at $s=0$. In particular, one
may substitute the Laplace transform $\tilde{f}(s)$ in equation (\ref{eq33}), 
and extract the asymptotic behaviour of ${\cal C}(t)$ at large $t$, by 
considering the limit of small $s$. The results we have obtained are 
simply a limiting case of those obtained by Weeks et al. \cite{Weeks96},
and we will therefore only quote the final result. 

One can show that at large $t$, ${\cal C}(t)$ has the following asymptotic 
behaviour
\begin{equation}
\label{eq37a}
{\cal C}(t)\sim\left\{
\mbox{
\begin{tabular}{c}   
$t^{4-\nu}$ if\; $2<\nu<3$\\
$t\ln t$  if\; $\nu=3$\\
$\frac{r}{\ave{\tau}}\,t$ if\;$\nu>3$
\end{tabular}
}
\right.,
\end{equation}
with $r=\frac{\nu-1}{\nu-3}>1$ being the motor's (finite) randomness 
coefficient if $\nu>3$.

The result obtained in the parameter region $2<\nu<3$, in which the
second moment of the distribution, or $r$, is formally infinite,
can be interpreted as a super-diffusive behaviour of the 
tracer particle. Such a behaviour is also present 
in the model studied by Kafri and 
co-workers \cite{Kafri04,Thanks_1_referee}.

We now consider the derivation of the Laplace transform of the
density-density correlation function
$S(q,t_1,t_2)=\langle\,e^{-iqd\,(N(t_1)-N(t_2))}\,\rangle$ in the long
time limit $t_1,t_2\rightarrow\infty$, which is given in equation
(\ref{GDeq4}).  Such a correlation function can be obtained from the
general expression for the probability-generating functional ${\cal Z}[j]$ by a
judicious choice of $g(u)$ and of the source function $j(x)$ in
equations (\ref{eq8},\ref{eq10}).  If we take $g(u)=\theta(u)$ and
\begin{equation}
j(x)=qd\,(\,\delta(x-t_2)-\delta(x-t_1)\,)\,
\label{eq38}
\end{equation}
then it is easy to see that ${\cal Z}[j]$ reduces to $S(q,t_1,t_2)$
for this particular choice of the functions $g(u)$ and $j(x)$. It also
follows from equation (\ref{eq21}) that $S(q,t_1,t_2)$ obeys the
following integral equation
\begin{equation}
 S(q,t_1,t_2)=1+(e^{-iqd}-1)\int_{0}^{t}\,d\zeta\,l(q,t_2+\zeta)\,,
\label{eq39}
\end{equation}
where $t=t_1-t_2$ and where $l(q,t_2+\zeta)$ is given by the
solution of the integral equation
\begin{equation}
l(q,t_2+\zeta)=l(t_2+\zeta)+(e^{-iqd}-1)\int_{0}^{\zeta}\,d\eta\,l(\zeta-\eta)\,l(q,t_2+\eta)  
\;\mbox{if}\;0\leq \zeta<t.
\label{eq40}
\end{equation}
This equation follows from (\ref{eq20}) for the particular choice we
made for $g(u)$ and $j(x)$. Such a system of equations can be solved
using Laplace transformation. Following Godreche and
Luck \cite{Godreche01}, we define the following Laplace transforms
\begin{eqnarray}
\tilde{l}(s\,;t_2)&=&\int_0^\infty\,d\zeta\,e^{-s\zeta}\,l(t_2+\zeta)\,,
\label{eq41}\\
\tilde{l}(q,s\,;t_2)&=&\int_0^\infty\,d\zeta\,e^{-s\zeta}\,l(q,t_2+\zeta)\,,
\label{eq42}\\
\tilde{S}(q,s\,;t_2)&=&\int_0^\infty\,dt\,e^{-st}\,S(q,t_2+t,t_2)\,.
\label{eq43}
\end{eqnarray}
Substituting equations (\ref{eq39},\ref{eq40}) in
(\ref{eq42},\ref{eq43}), one obtains, after some trivial
manipulations, the following result for  $\tilde{S}(q,s;t_2)$ 
\begin{equation}
\tilde{S}(q,s\,;t_2)=\frac{1}{s}\,
\left(1+\frac{e^{-iqd}-1}{1+(1-e^{-iqd})\,\tilde{l}(s)}\,\tilde{l}(s\,;t_2)\,
\right)\,
\label{eq44}
\end{equation}
where $\tilde{l}(s)$ is the Laplace transform of $l(u)$, given by
(\ref{eq6A}). In the long time limit $t_2\rightarrow \infty$,
$l(t_2+\zeta)\rightarrow \frac{1}{\ave{\tau}}$ and $\tilde{l}(s\,;t_2)=
\frac{1}{\ave{\tau}s}$, which is independent of $t_2$. Therefore,
$\tilde{S}(q,s\,;t_2)$ is also independent of $t_2$ and we write it
simply as $\tilde{S}(q,s)$ in this limit. Substituting $\tilde{l}(s)$
by its expression in terms of $\tilde{f}(s)$, as given by equation
(\ref{eq6A}), in equation (\ref{eq44}), we conclude that
$\tilde{S}(q,s)$ is given by equation (\ref{GDeq4}) \cite{fnote3}.

We will now consider the relation between the Fourier and the Laplace
transforms of $S(q,t)$. From its definition, one immediately concludes
that for $t>0$,
\begin{equation}
S(q,-t)=S(-q,t)=\overline{S(q,t)}\,.
\label{eq45}
\end{equation}
If we substitute this identity in the definition of $S(q,\omega)$, we
obtain
\begin{eqnarray}
S(q,\omega)&=&\int_{-\infty}^0\,dt\,e^{i\omega t}\,S(q,t)+
\int_0^{\infty}\,dt\,e^{i\omega t}\,S(q,t)\nonumber\\
&=&\overline{\int_0^\infty\,dt\,e^{i\omega t}\,S(q,t)}+
\int_0^{\infty}\,dt\,e^{i\omega t}\,S(q,t)\nonumber\\
&=&2\,Re\,\tilde{S}(q,s=-i\omega)\,,
\label{eq46}
\end{eqnarray}
where we have performed the change of variable $t\rightarrow -t$ in
the first term of the first line of (\ref{eq46}). Equation
(\ref{GDeq5}) then follows from (\ref{GDeq4}) and from (\ref{eq46}).
Finally, one obtains (\ref{GDeq6}) by substituting
$\tilde{f}(-i\omega)= (1-i\omega\tau)^{-1}$, valid for a simple
Poisson process, in (\ref{GDeq5}).

Before closing this section, we will briefly explain how one can
obtain the expression for the spin-spin correlation function
$\langle\vec{\sigma}_{t_1}\cdot\vec{\sigma}_{t_2}\rangle$ in the time
domain in the cases in which the waiting-time distribution of the
renewal process is given either by a single Poisson process, as in
(\ref{GDeq8}), or in which the waiting time-distribution is given by
the convolution of two Poisson processes with the same characteristic
time $\tau/2$, as in (\ref{GDeq9}), leaving the details to \ref{apC}.  
If the waiting-time distribution $f(\zeta)$ is a simple
exponential, then, substituting the result for its Laplace transform
$\tilde{f}(s)$ in equation (\ref{GDeq7}), with $Q=3$, one observes that the
Laplace transform of the spin-spin correlation function has a simple
pole, and one can directly read
$\langle\vec{\sigma}_{t_1}\cdot\vec{\sigma}_{t_2}\rangle$ from it.  If
the waiting-time distribution is given by the convolution of two
Poisson processes with the same characteristic time, the Laplace
transform of
$\langle\vec{\sigma}_{t_1}\cdot\vec{\sigma}_{t_2}\rangle$, as given by
equation (\ref{GDeq7}), with $Q=3$, has two simple poles and this
expression can always be written in terms of partial fractions
involving one or the other of these poles. In that case, one can also
read $\langle\vec{\sigma}_{t_1}\cdot\vec{\sigma}_{t_2}\rangle$
directly from it.
\section{Conclusion and outlook}
\label{secV}
We have modelled the dynamics of a processive or rotary molecular
motor as a renewal process, in line with the work of Svoboda and
co-wokers. Using a functional technique, we have computed the
mean-square deviation of the distance travelled by a processive motor
and extracted its asymptotic limit at large times. For renewal
processes composed of two or three Poisson substeps, we have computed
this function outside the asymptotic regime, given the relevance of
such processes for the study of myosin V and kinesin. It follows from
our results that the measurement of this correlation function would
permit one to extract additional information concerning the time constants
which characterise the motor's chemical cycle.

We have also used the same functional method to compute the
density-density correlation function of an ensemble of independent
processive motors, which can be measured using the experimental
techniques developed by Cappello and co-workers. We have also shown
that in a particular limit such a function reduces to the spin-spin
correlation function of a 'random-clock' model that has applications
to the dynamics of rotary motors, such as F1-ATPase. The measurement
of this correlation function, followed by its Fourier transformation
to frequency space, would permit one, in both cases, to fit such a quantity
to the theoretical results discussed above. Thus, one could obtain all the
time constants characterising the molecular motor's chemical cycle,
even in the case of chemical cycles composed of a large number of
rate-limiting substeps.

As possible avenues of future research, one can indicate at least
two experimental issues that still 
need to be addressed. The first experimental issue is the repetition of the 
experiments of Cappello and co-workers with an interference mask
with a smaller period or the use of myosin V, rather than kinesin,
as the subject of study of such an experiment, which would allow 
the measurement of the density-density correlation function
away from the limit of long wavelengths, where the motor simply behaves
as a Brownian particle. The second issue is the direct
measurement of the spin-spin correlation function in F1-ATPase using
fluorescence microscopy and how one can improve the time resolution of the
present technique. 

In closing, we may say that the results presented in this paper
can be used, in the different experimental contexts to which they
apply, to provide for a precise fitting of the time constants
associated with the rate-limiting steps of a molecular motor chemical
cycle. It remains to be seen to what extent such
information is of crucial importance to the understanding of the
chemical kinetics of processive or rotary molecular motors, or if a
qualitative understanding of the nature of the chemical cycle is by
itself sufficient. \\

{\bf Acknowledgments:} We acknowledge many helpful discussions with
P.  Pierobon, M.  Badoual, A. Vilfan, G. Lattanzi, P. Benetatos, K.
Kroy, R. Thul and G. Cappello.  It is also a pleasure to acknowledge
the referees of Physical Biology for their pertinent questions and
comments, and the Style Editor of the journal, Valerie Parsegian,
for her help with the subtleties of the English language in general
and the improvement of the presentation in particular. 
J.E.S. and E.F. acknowledge financial support from
the DFG in the framework of the Sonderforschungsbereich SFB 413/TP C6
and from IBM Deutschland in the framework of FZ J\"ulich
TTB/1030.03.03 contract.  A.P. acknowledges financial support from the
Marie-Curie Fellowship no.  HPMF-CT-2002-01529 and from the "Aides
Jeunes Chercheurs" of the University of Montpellier 2, France.

\newpage

\section*{Glossary}
{\bf Bead-motor assay:} Experimental apparatus where a spherical plastic bead 
of micrometric size is coupled to a single molecular motor. The position of
the bead can be controlled with high precision using two intense laser beams
focused at a spot (known as optical tweezers). Due to its optical properties,
the bead is attracted to the focus of the two beams and may be controlled at
a distance.
\\
{\bf Molecular motor stepping:} Directed motion of a processive molecular
motor (e.g. myosin V, kinesin, dynein) along a specific molecular track
in the cell (e.g. actin in the case of myosin V, tubulin in the case of
kinesin and dynein). Such a motion is composed of individual steps of fixed
length, which occur at random times, determined by the chemistry of the 
process.
\\
{\bf Probability-generating functional:} Mathematical object which encodes
in itself all the information that can be obtained (measured) from a random
process. In the main text, we have computed the explicit form of such an
object (\ref{eq25}) for a renewal process where the interval distribution function was
a simple exponential. From such an expression,
one can obtain all the correlation function pertaining to such a renewal
process. Thus, the ultimate goal of applied probability theory is the
computation of the probability-generating functional for a given random process. All too
often, one has to content oneself with well less than that. 
\\
{\bf Renewal process:} Random counting process where an integer variable 
is increased by one unit at random times, the statistical
distribution of the length of time intervals between sucessive counting events being 
a known function.

\appendix

\section{Mean-square deviation of the displacement
  for a renewal process composed of two or three Poisson substeps}
\label{apA}
In this appendix, we derive the explicit form, valid for arbitrary
time $t$, of the mean-square deviation of the distance travelled by
the motor ${\cal C}(t)$, for a renewal process whose waiting-time
distribution is given by the convolution of two or three Poisson
processes. As can been seen from equation (\ref{eq33}), the knowledge
of the renewal function $l(u)$ suffices to determine ${\cal C}(t)$.

We start with the simplest case, namely the case in which the
waiting-time distribution is given by the convolution of two Poisson
processes. In that case, the Laplace transform of the waiting-time
distribution is given by
$\tilde{f}(s)=\frac{1}{(1+s\tau_1)(1+s\tau_2)}$, as follows from
equation (\ref{eq37}). Substituting this result in equation
(\ref{eq6A}), one can show that the Laplace transform of the renewal
function can be written in terms of partial fractions as
\begin{equation}
\label{apeq1}
\tilde{l}(s)=\frac{1}{\tau_1+\tau_2}\,\left(\frac{1}{s}
-\frac{1}{s+1/\tau_1+1/\tau_2}\right)\,.
\end{equation}
From this equation, one can directly determine the inverse transform
$l(u)$, which is given by
\begin{equation}
l(u)=\frac{1-e^{-\left(\frac{1}{\tau_1}+\frac{1}{\tau_2}\right)u}}{\tau_1+\tau_2}\,,
\label{apeq2}
\end{equation}
which tends to $l(u)\rightarrow \frac{1}{\tau_1+\tau_2}$ in the limit
of large $u$, in agreement with Blackwell's renewal theorem.
Substituting this result for $l(u)$ in equation (\ref{eq33}) and performing
the integration over $u$, one obtains for ${\cal C}(t)$ the result
\begin{equation}
{\cal C}(t)=
\frac{r\,t}{\tau_1+\tau_2}+\frac{2\tau_1^2\tau_2^2}{(\tau_1+\tau_2)^4}
\left(1-e^{-(\frac{1}{\tau_1}+\frac{1}{\tau_2})t}\right)\,,
\label{apeq3}
\end{equation}
where $r=\frac{\tau_1^2+\tau_2^2}{(\tau_1+\tau_2)^2}$ is the randomness
parameter for a motor whose chemical cycle
is composed of two rate-limiting steps. 
It can be easily checked that this function has the correct asymptotic
forms at small and large $t$, has given by (\ref{GDeq3}).
We have plotted
this function in figure \ref{F1}, with $\tau_1=\tau_2=1/2$, i.e.
when $r=1/2$.

In the case of a renewal processes whose waiting-time distribution is
given by the convolution of three Poisson processes, it follows from
(\ref{eq37}) that $\tilde{f}(s)=\frac{1}{(1+s\tau_1)(1+s\tau_2)(1+s\tau_3)}$.
Substituting this result in equation (\ref{eq6A}), one can show
that the Laplace transform of the renewal function can be written as
\begin{equation}
\label{apeq4}
\!\!\!\!\!\!\!\!\!\!\!\!\!\!\!\!
\tilde{l}(s)=\frac{1}{\tau_1+\tau_2+\tau_3}\,\left[
\frac{\gamma_+}{\gamma_+-\gamma_-}\left(\frac{1}{s}
-\frac{1}{s+\gamma_-}\right)
-\frac{\gamma_-}{\gamma_+-\gamma_-}\left(\frac{1}{s}
-\frac{1}{s+\gamma_+}\right)\right]\,,
\end{equation}
where $\gamma_{\pm}=\frac{1}{2}\left[\left(\frac{1}{\tau_1}+
\frac{1}{\tau_2}+\frac{1}{\tau_3}\right) \pm\sqrt{\left(\frac{1}{\tau_1}+
\frac{1}{\tau_2}+\frac{1}{\tau_3}\right)^2-4\left(\frac{1}{\tau_1\tau_2}+
\frac{1}{\tau_1\tau_3}+\frac{1}{\tau_2\tau_3}\right)}\right]$.
One can also directly determine the inverse transform $l(u)$ from
this equation, which is given by
\begin{equation}
l(u)=\frac{1}{\tau_1+\tau_2+\tau_3}\,\left(\frac{\gamma_+}{\gamma_+-\gamma_-}\,
(1-e^{-\gamma_-u})-
\frac{\gamma_-}{\gamma_+-\gamma_-}\,(1-e^{-\gamma_+u})\right)\,.
\label{apeq5}
\end{equation}
This function tends to $l(u)\rightarrow
\frac{1}{\tau_1+\tau_2+\tau_3}$ in the limit of large $u$, which also
agrees with Blackwell's renewal theorem.  Substituting this result for
$l(u)$ in equation (\ref{eq33}) and performing the integration over
$u$, one obtains for ${\cal C}(t)$ the result
\begin{eqnarray}
{\cal C}(t)&=&\frac{
  rt}{\tau_1+\tau_2+\tau_3}+
\frac{2}{(\tau_1+\tau_2+\tau_3)^2}\left(
\frac{\gamma_+}{\gamma_-^2(\gamma_+-\gamma_-)}\left(1-e^{-\gamma_-t}\right)\right.\nonumber\\
&&\left.\mbox{}-\frac{\gamma_-}{\gamma_+^2(\gamma_+-\gamma_-)}\left(1-e^{-\gamma_+t}\right)\right),
\label{apeq6}
\end{eqnarray}
where $r=\frac{\tau_1^2+\tau_2^2+\tau_3^2} {(\tau_1+\tau_2+\tau_3)^2}$
is the randomness parameter for a motor whose chemical cycle is
composed of three rate-limiting steps.  It can be easily checked that
this function has the correct asymptotic forms at small and large $t$,
has given by (\ref{GDeq3}).  We have plotted this function in figure
\ref{F2}, with $\tau_1=\tau_2=\tau_3=1/3$, i.e. when $r=1/3$.
\section{Density-density correlation function in Fourier space
for a renewal process composed of two or three Poisson substeps}
\label{apB}
All one has to do is to substitute the appropriate form for
$\tilde{f}(-i\omega)$ in equation (\ref{GDeq5}). For a renewal process
whose waiting-time distribution is given by the convolution of two
Poisson processes,
$\tilde{f}(-i\omega)=\frac{1}{(1-i\omega\tau_1)(1-i\omega\tau_2)}$ and
$S(q,\omega)$ is given by
\begin{equation}
\label{apBeq1}
S(q,\omega) = \frac{2(\tau_1+\tau_2)\,(1-\cos(qd))\,
(r+\alpha\omega^2)}{(\omega(\tau_1+\tau_2)-\sin(qd))^2+
(1-\cos(qd)-\tau_1\tau_2\,\omega^2)^2}\,,
\end{equation}
where $r=\frac{\tau_1^2+\tau_2^2}{(\tau_1+\tau_2)^2}$ and 
$\alpha=\frac{\tau_1^2\tau_2^2}{(\tau_1+\tau_2)^2}$.

For a renewal process whose waiting-time distribution is given by the
convolution of three Poisson processes,
$\tilde{f}(-i\omega)=\frac{1}{(1-i\omega\tau_1)(1-i\omega\tau_2)
  (1-i\omega\tau_3)}$ and $S(q,\omega)$ is given by
\begin{equation}
\label{apBeq2}
S(q,\omega)=
{2(\tau_1+\tau_2+\tau_3)\,(1-\cos(qd))\, (r+\alpha\omega^2+\beta\omega^4)}/N
\end{equation}
where 
\begin{eqnarray}
N &=& [\omega(\tau_1+\tau_2+\tau_3)-\sin(qd)
  -\tau_1\tau_2\tau_3\,\omega^3]^2
  \nonumber \\ \quad &&+
[1-\cos(qd)-(\tau_1\tau_2+\tau_1\tau_3+\tau_2\tau_3)\,\omega^2]^2 \,,
\end{eqnarray}
and $r=\frac{\tau_1^2+\tau_2^2+\tau_3^2}
{(\tau_1+\tau_2+\tau_3)^2}$, $\alpha=\frac{\tau_{1}^2\tau_{2}^2
+\tau_{1}^2\tau_{3}^2+\tau_{2}^2\tau_{3}^2}{(\tau_1+\tau_2+\tau_3)^2}$ and
$\beta=\frac{\tau_{1}^2\tau_{2}^2\tau_{3}^2}{(\tau_1+\tau_2+\tau_3)^2}$.
\section{Spin-spin correlation function for a renewal
  process composed of two Poisson substeps}
\label{apC}
We first obtain the spin-spin correlation function
in the time domain, as given by equation (\ref{GDeq9}), if the
two time constants are equal, i.e. if $\tau_1=\tau_2=\tau/2$.
In such a case, the Laplace transform of the waiting-time
distribution is given by  $\tilde{f}(s)=\frac{1}{(1+s\tau/2)^2}$.
Substituting such a result in equation (\ref{GDeq7}), 
one can write the Laplace transform of the spin-spin correlation
function in terms of partial fractions as
\begin{equation}
\label{apCeq1}
{\cal L}\,\langle \vec{\sigma}_{t_1}\cdot\vec{\sigma}_{t_2}\rangle =
Re\,\left[
\frac{3}{4}\,\frac{1}{s+\frac{2}{\tau}(1-e^{-i\pi/3})}
+\frac{1}{4}\,\frac{1}{s+\frac{2}{\tau}(1+e^{-i\pi/3})}\right]\,.
\end{equation}
From such a result, one can immediatelly read its inverse Laplace
transform, which is given by equation (\ref{GDeq9}). Note that
if the renewal process is a simple Poisson
process, $\tilde{f}(s)=\frac{1}{1+s\tau}$ and the Laplace transform
of the spin-spin correlation function involves a simple pole, which
make its inversion trivial and we obtain (\ref{GDeq8}).

For a general choice of the time constants $\tau_1$ and $\tau_2$, one
can also write the Laplace transform of the spin-spin correlation
function in terms of partial fractions, which permits to invert such a
transform. However, extracting the real part of such an expression
becomes a cumbersome exercise, albeit a trivial one. It is simpler,
also from the viewpoint of the experimental fitting of the data, to
consider instead the Fourier transform of the spin-spin correlation
function, which is defined as $FT \langle
\vec{\sigma}_{t_1}\cdot\vec{\sigma}_{t_2}\rangle =
\int_{-\infty}^{+\infty}\,dt\,e^{i\omega t}\,\langle
\vec{\sigma}_{t_1}\cdot\vec{\sigma}_{t_2}\rangle$, with $t=t_1-t_2$.
Since one can write the spin-spin correlation function as $\langle
\vec{\sigma}_{t_1}\cdot\vec{\sigma}_{t_2}\rangle=
\frac{1}{2}(\,S(2\pi/3,t)+S(2\pi/3,-t)\,)$ (with $d=1$), 
the Fourier transform of this function is given by
\begin{equation}
\label{apCeq2}
FT \langle \vec{\sigma}_{t_1}\cdot\vec{\sigma}_{t_2}\rangle =\frac{1}{2}\,(
S(2\pi/3,\omega)+S(2\pi/3,-\omega))\,.
\end{equation}
where $S(q=2\pi/3,\omega)$ is given by (\ref{apBeq1}), valid for
a renewal process composed of two Poisson substeps. We finally
obtain
\begin{eqnarray}
\label{apCeq3}
\!\!\!\!\!\!\!\!\!\!\!\!\!\!\!\!
FT \langle \vec{\sigma}_{t_1}\cdot\vec{\sigma}_{t_2}\rangle &=&
\frac{3}{2}(\tau_1+\tau_2)(r+\alpha\omega^2)\,
\left[\,\frac{1}{(\omega(\tau_1+\tau_2)-\sqrt{3}/2)^2+
(3/2-\tau_1\tau_2\,\omega^2)^2}\right.\nonumber\\
&&\mbox{}+\left.
\frac{1}{(\omega(\tau_1+\tau_2)+\sqrt{3}/2)^2+
(3/2-\tau_1\tau_2\,\omega^2)^2}\,\right]\,,
\end{eqnarray}
where $r=\frac{\tau_1^2+\tau_2^2}{(\tau_1+\tau_2)^2}$ and
$\alpha=\frac{\tau_1^2\tau_2^2}{(\tau_1+\tau_2)^2}$.  We plotted this
function for three different choices of the time constants $\tau_1$
and $\tau_2$, such that $\tau_1+\tau_2=1$, in figure \ref{F4}.


\begin{thebibliography}{10}
\bibitem{fnote0}
The best analogy one can make between a molecular motor and a macroscopic
  thermodynamic engine would be to a fuel-cell coupled to an electric engine,
  operating at constant temperature. However, such an analogy is far from being
  perfect, since in such an arrangement the generation of an electric current
  and its convertion into motion occur in physically distinct regions, whereas
  in a molecular motor the conformational changes of the molecule, which
  engender its overall motion, occur directly due to chemical transformations.
\bibitem{fnote_Ali}
In reality, the step-size of $34.8$ nm is just an average value, since myosin V
  may perform steps involving the attachment of the free head to the 11th or
  13th unit (with respect to that of the bound head) along the actin
  spiral\cite{Ali02}. The description of the motion of myosin V in terms of a
  renewal process with a constant step size is thus just an approximation, but
  which should be valid for length scales (times) much larger than the distance
  of 5.5 nm (crossing-time) between the neighbouring units.
\bibitem{fnotePeclet}
The randomness coefficient is nothing more than the inverse of the Peclet
  number $\propto dv/D$, used in the study of polydisperse systems, where $d$
  are the typical dimensions of the object, $v$ its average velocity and $D$
  its diffusion constant.
\bibitem{fnote_pedantic}
When we say that a renewal process is composed of ${\cal M}$ rate-limiting
  processes, what we of course mean is that the waiting-time distribution
  $f(\zeta)$ of such a process is given by the convolution of ${\cal M}$ simple
  Poisson processes, with characteristic times $\tau_1,\cdots,\tau_{\cal M}$,
  in line with the model introduced by Svoboda and coworkers. We believe that
  the number of words spared in the text justifies commiting such a small
  imprecision.
\bibitem{fnote1}
The randomness parameter can be larger than one, e. g. if one allows for the
  presence of thermal noise, through the Brownian motion of the dielectric
  bead, or if the motor can perform backward steps. The first effect can be
  accounted for if one assumes that such noise is uncorrelated with the renewal
  process ocurring in the motor\cite{Svoboda94,Badoual04} and contributes
  additively to the motion of the bead. The second effect is only pronounced if
  one is probing the motor close to the stall force or working at very low ATP
  concentration.
\bibitem{fnote2}
The turnover time can also be determined by the measurement of ${\cal C}(t)$ at
  small times $t$, but such a measurement is likely to be much less precise
  than the measurement of $\ave{\tau}$ obtained from the average displacement
  of the motor, which extends itself through several decades.
\bibitem{fnote_CB}
Strictly speaking, the quantity measured in the experiments of Cappello and
  co-workers\cite{Cappello03,Badoual04} is the intensity-intensity correlation
  function of the light scattered by the bead, which is given by the real part
  of $S(q,t)$, with $q=2\pi/\Lambda$, where $\Lambda$ is the length of the mask
  used to create the interference pattern. Therefore, one has that the Fourier
  transform of $Re \;S(q,t)$ is in fact given by
  $\frac{1}{2}(S(q,\omega)+S(q,-\omega))$, as in the case of the spin-spin
  correlation function, see below in the text.
\bibitem{fnote_Cappello_Exp}
Since $S(q,\omega)$ reduces to equation (\ref{GDeq6a}) at small $q$ and
  $\omega$, it is advisable that experiments such as those carried by Cappello
  and co-workers \cite{Cappello03,Badoual04}, in which the wave-vector
  $q=2\pi/\Lambda$ is kept constant throughout the experiment, are performed at
  high values of $q$, or small values of the mask length $\Lambda$. Otherwise,
  if $qd\ll 1$, one will not be able to determine more than the two parameters
  $v$ and $D$, which characterise the motor's Brownian motion in the large
  wavelength limit. In the experiments reported in reference \cite{Cappello03},
  the mask length $\Lambda=190$ nm, whereas $d$, the step size of kinesin, the
  motor whose motion was studied, is equal to $8$ nm. One thus has that
  $qd\approx 0.265$, i.e. one is still in the limit of low $q$ and one will not
  be able to fully determine the characteristic times pertaining to the three
  or four rate-limiting steps of the chemical cycle of kinesin. Since for
  myosin V, $d=35$ nm, such experiments will probably yield clearer results if
  performed with this molecular motor.
\bibitem{pol_signs}
For the three different levels of polarisation measured by Adachi and
  co-workers\cite{Adachi00}, the signs in the expressions for $(h(T), v(T))$,
  should be, for levels a, b, c, respectively (1,1), (-1,1), (-1,-1).
\bibitem{Thanks_2_referee}
We would explicitly like to thank the second referee of Physical Biology for
  having pointed out the possibility to carry out the measurement of the
  spin-spin correlation function using this method.
\bibitem{Thanks_1_referee}
We would explicitly like to thank the first referee of Physical Biology for
  having raised the issue of a waiting-time distribution with fat tails.
\bibitem{fnote3}
One can obtain equation (\ref{GDeq4}) directly, without having to consider the
  long time limit $t_2\rightarrow \infty$, if one chooses the waiting-time
  distribution for the first step to be given not by $f(\zeta)$, but by ${\cal
  F}(\zeta)
  =\frac{1}{\ave{\tau}}\,\int_{\zeta}^{\infty}\,d\eta\,f(\eta)$\cite{Macchi71}.
  Such an initial distribution completely kills off all the transients and
  garanties that the time-translation invariant regime is valid right after
  $t_2=0$. In fact, the choice of such a distribution can reduce the
  computation of multiple-time correlations functions such as ${\cal C}(t)$ to
  the computation of single-time correlation functions. However, in a real
  experiment performed with a molecular motor, there is no guaranty that the
  waiting-time distribution for the first step is actually given by ${\cal
  F}(\zeta)$ and therefore it is crucial that the measurements, if they are not
  to be influenced by initial transients which may even vary from one
  experiment to the next, are taken in the limit $t_1,t_2\gg \ave{\tau}$, i.e.
  after the motor has completed several steps (chemical cycles). We have
  therefored chosen not to consider this particular case in our discussion.
\end{thebibliography}
\end{document}